\PassOptionsToPackage{table,xcdraw}{xcolor}
\documentclass{article}    %
\usepackage{arxiv}
\usepackage{bookmark}
\AtBeginDocument{%
	\providecommand\BibTeX{{%
			\normalfont B\kern-0.5em{\scshape i\kern-0.25em b}\kern-0.8em\TeX}}}

\usepackage[numbers]{natbib}
\usepackage{amsmath}
\usepackage{amssymb}
\usepackage{multicol}
\usepackage{algorithmic}
\usepackage[ruled,vlined]{algorithm2e}
\usepackage{listings}
\usepackage{graphicx}
\usepackage{textcomp}
\usepackage[]{footmisc}
\usepackage{xfp}
\usepackage{stfloats}
\usepackage{booktabs}
\PassOptionsToPackage{hyphens,spaces,obeyspaces}{url}\usepackage{hyperref} 
\usepackage{pdflscape}
\usepackage{longtable}
\makeatletter
\def\LT@makecaption#1#2#3{%
	\LT@mcol\LT@cols c{\hbox to\z@{\hss\parbox[t]\LTcapwidth{%
				\footnotesize\bgroup\par\centering\@IEEEtabletopskipstrut{\normalfont\footnotesize #2}\\{\normalfont\footnotesize\scshape #3}\par\addvspace{0.5\baselineskip}\egroup\endgraf%
				\@IEEEtablecaptionsepspace}%
			\hss}}}
\makeatother
\usepackage[flushleft]{threeparttable}
\usepackage{threeparttablex}
\usepackage{setspace}
\usepackage{tikz}
\usepackage[most]{tcolorbox}
\usepackage{flowchart}
\usepackage{multirow}
\usepackage{lscape}
\usepackage{array}
\usepackage{colortbl}
\usepackage{ragged2e}
\usepackage{amsfonts}
\usepackage{subcaption}
\usepackage{cleveref}
\usepackage{adjustbox}
\crefname{section}{§}{§§}
\Crefname{section}{§}{§§}
\newcommand\tab[1][1cm]{\hspace*{#1}}
\usepackage{tikz}
\usepackage{pgfplots}
\usetikzlibrary{shapes.geometric, arrows, trees}
\usepgfplotslibrary{external}
\pgfplotsset{compat=1.17}
\usepackage{pifont}%
\newcommand{\cmark}[1][2.5]{\tab[#1pt]\ding{51}}%
\definecolor{Gray}{gray}{0.9}
\definecolor{lightgray}{gray}{0.95}
\newcolumntype{L}[1]{>{\centering\arraybackslash\columncolor{Gray}}m{#1}}
\newcolumntype{g}[1]{>{\raggedright\arraybackslash\columncolor{Gray}}m{#1}}
\newcolumntype{f}[1]{>{\raggedleft\arraybackslash\columncolor{Gray}}m{#1}}
\newcolumntype{C}[1]{>{\columncolor{Gray}}p{#1}}
\colorlet{tablerowcolor}{white} %
\newcommand{\rowcol}{\rowcolor{tablerowcolor}} %
\newlength{\maxheadlen}
\newlength{\maxsubheadlen}

\newcommand*{\subhead}[2]{
	\multicolumn{1}{p{#1cm}}{\multirow{6.5}{*}{\tab[\fpeval{(#1-0.25)*0.5}cm]\rotatebox{60}{\makebox[\maxsubheadlen][l]{#2}}}}
}
\newlength{\maxsubrotheadlen}
\newcommand*{\subrothead}[3][60]{
	\multirow{#2}{*}{\rotatebox{#1}{\makebox[\maxsubrotheadlen][l]{#3}}}
}
\newlength{\maxrotheadlen}
\newcommand{\rothead}[3][60]{\multirow{#2}{*}{\rotatebox{#1}{\makebox[\maxrotheadlen][l]{#3}}}}

\newcommand{\pie}[1]{%
\begin{tikzpicture}
 \draw (0,0) circle (0.75ex);\fill (0,0) -- (0,0.75ex) arc (90:90-#1:0.75ex) -- cycle; %
\end{tikzpicture}%
}
\newcommand{\Circle}{
	\begin{tikzpicture}
	\draw (0,0) circle (0.75ex);
	\end{tikzpicture}%
}
\newcommand{\RIGHTcircle}{\pie{180}}
\newcommand{\CIRCLE}{\pie{360}}

\title{Identifying Authorship Style in Malicious Binaries: Techniques, Challenges \& Datasets}

\author{
	Jason Gray\thanks{Royal Holloway University of London} , 
	Daniele Sgandurra\thanks{Royal Holloway University of London} , and
	Lorenzo Cavallaro\thanks{King's College London} 
}
\date{}
\renewcommand{\headeright}{}    
\renewcommand{\undertitle}{}    

\begin{document}
	\maketitle

\begin{abstract}
	Attributing a piece of malware to its creator typically requires threat intelligence. Binary attribution increases the level of difficulty as it mostly relies upon the ability to disassemble binaries to identify authorship style.  Our survey explores malicious author style and the adversarial techniques used by them to remain anonymous. We examine the adversarial impact on the state-of-the-art methods. We identify key findings and explore the open research challenges. To mitigate the lack of ground truth datasets in this domain, we publish alongside this survey the largest and most diverse meta-information dataset of 15,660 malware labeled to 164 threat actor groups.

\end{abstract}


\thispagestyle{plain}
\pagestyle{plain}

\lstset{numbers=left, numberstyle=\tiny, stepnumber=1, numbersep=5pt}
\lstset{language=Python}

\section{Introduction}\label{sec:introduction}

Malicious software (malware) remains one of the biggest threats to organizations, and there seems no sign of this changing in the near future \cite{Symantec2019}. Identifying malware authors to a person, group or country provides evidence to analysts of the wider goals of threat actors. Furthermore, it provides a method to counter cyber attacks and disrupt the malware economy through public indictment \cite{RidBuchanan2015,odni2018}.

The current and only method for authorship attribution used by analysts involves prolonged analysis of the threat actor over a long duration and within different phases of the killchain \cite{LockheedMartin2015}. Part of this process includes gathering features such as network analysis and exploitation techniques referred to as \emph{indicators of compromise} as well as relying on known databases of Tactics, Techniques and Procedures (TTPs). %

Sometimes there exists no wider context, especially if the threat actor is unknown to the victim. In very few cases, analysts discover the malware source code and use this to determine attribution through source code authorship attribution \cite{Frantzeskouetal2007,KrsulSpafford1997,Caliskanetal2015,Laytonetal2010}. However, released source code  leads to copycat attacks or the malware no longer used \cite{Gamer2016}. This means defenders often find themselves with only the malware binary as evidence. The quicker the defenders analyse the malware and identify a probable threat actor, the quicker they can understand and contextualize an attack (including if they must contact an authority and which relevant authority), which leads to a quicker response time and attack mitigation. 

The specific problem of identifying an author of piece of malware is known as Malware Authorship Attribution (MAA). However, using the binary alone represents a difficult problem due to the complexities of program provenance \cite{Rosenblumetal2010}. Despite this, the binary still provides interesting artifacts on author style, \textit{e.g.,} implementation of encryption, propagation, mutation or even the setup of command and communication servers within the malware infrastructure. Even though the demand for malware attribution continues to increase, we notice few publications detailing the methods of malware authorship attribution.

Recent work \cite{Tempesttetal2018, Burrowsetal2014, Alrabaeeetal2017, Kalgutkaretal2019, Brennanetal2012} informed the wider authorship attribution field. \citet{Tempesttetal2018} wrote a survey on the wider topic of stylometry focusing on de-anonymizing text. The survey by \citet{Burrowsetal2014} focuses on the attribution of source code up until 2011 and highlights the positive use of Machine Learning techniques in the authorship attribution field. \citet{Brennanetal2012} introduce the notion of exploring the adversarial approach toward the stylometry problem and provide novel datasets to aide this research direction. \citet{Kalgutkaretal2019} provide further insight on the code authorship attribution problem by exploring the use of features in benign source and binary code attribution as well as the attribution models and methods. They also present the challenges in the research field and incorporate the field of plagiarism detection. Finally, \citet{Alrabaeeetal2017} discuss three state-of-the-art techniques for the single-author binary authorship problem \cite{Caliskanetal2018,Alrabaeeetal2014,Rosenblumetal2011} and provide promising results from applying malware to these systems.

The current state-of-the-art systems show promising results on attributing programs where author style remains unaltered apart from compilation techniques. However, there exist few attempts to extend these systems to consider author masking techniques such as those used by some Advanced Persistent Threat (APT) groups). This limitation to the current state-of-the-art systems opens them to attack and thus there exists the need to fully understand the adversarial challenges to authorship attribution of malware. 

\paragraph{Contributions} Our contributions include a thorough systematization of the malware authorship attribution problem focusing on the data modeling techniques, datasets and features used for attribution to allow the community to understand how each paper builds upon each other and the shortcomings within the current research. We review eighteen attribution systems. We compare them in terms of techniques, features, efficacy, functionality and adversarial robustness. 

We discover there exist only two publicly available author-labeled malware datasets, both of which contain significant flaws such as non unique labels. Furthermore, we found the current features used for author style remain varied with no clear consensus on authorship style (42 of a total of 72 features were used separately by research groups). The current state-of-the-art systems remains inapplicable to real world use cases. The majority of systems fail to take into account modern malware development methods, \textit{e.g.,} assuming multiple authors. Additionally, researchers use non-representative datasets of the real world which introduce adversarial issues surrounding open world assumptions, continuous learning, concept drift and obfuscation. On top of this, the majority of attribution systems from existing research lack reproducibility owing to system unavailability, systems no longer working, or the literature omitting fundamental details.

Focusing on the dataset problem, we contribute by publishing a labeled meta-information dataset of 15,660 malware. We extensively use open-source intelligence to build a list of APT groups and then gather hashes of malware to which we verify their legitimacy against VirusTotal. We use Natural Language Processing techniques to gather the most high likely label for a hash from various open-source intelligence material. This dataset is the largest verified APT labeled malware dataset to date. We searched 896 files made up of a mixture of PDFs, CSVs, rules, and indicator of compromise files. We found 15,660 unique hashes which we have labeled to 164 APT groups. Furthermore, we identified an additional 7,485 unique hashes. For these unlabeled hashes, we record the top 5 keywords from the file and the keywords of the metadata.

Our work complements \citet{Kalgutkaretal2019} and \citet{Brennanetal2012} by extending the application of authorship attribution to malware by including a full detailed analysis of malware author style, features and adversarial approach. We expand upon the work by \citet{Alrabaeeetal2017} and incorporate the multiple author attribution problem into the conversation of MAA. We also note the survey by \citet{Xueetal2019} which focuses on general Machine Learning based program analysis whereas we focus purely on authorship style gained from program analysis using multiple data modeling techniques.

Section \ref{sec:background} presents the background to threat actors, authorship attribution and adversarial techniques. Section \ref{sec:systemization} systematizes the MAA problem, focusing on the data modeling techniques, authorship style, features and datasets. Section \ref{sec:discussion} discusses real-world application of the current state-of-the-art, looking at the challenges and recommendations for future work. Finally, in Section \ref{sec:new_dataset}, we present the method we used to create a new APT malware dataset for the research community.

\section{Background: Threats Actors, Authorship Attribution, \& Adversarial Techniques}\label{sec:background}
In this section, we set out the background to the \textit{malware authorship attribution (MAA)} problem. MAA is the identification of the author of an unknown malicious file\footnote{Throughout the paper, we refer to malicious files as ``malware'', ``malware binaries'' or ``malicious binaries'' interchangeably.}. In particular, we define the authors we wish to identify as  \textit{Threat Actors}. We also explore the more wider form of the MAA problem and consider the types of adversary attacks which MAA systems are likely to face.

\subsection{Threat Actors}\label{sec:apt-background}

Although the threat actors with the greater skill level tend to use better adversarial techniques \cite{BartholomewGuerrero-Saade2016}, they also tend to possess unique styles when using custom-made tools. Naturally, if any attackers use commercially available or open source tools, then the author of the tool is not necessarily the threat actor. 
As we wish to focus on identifying author style of threat actors, we shall look to focus on where style exists \textit{i.e.} within custom tools. These tools are generally produced by Advanced Persistent Threats (APTs).

\paragraph{Advanced Persistent Threats}
APTs represent the most sophisticated attackers. The US National Institute of Standards and Technology (NIST) provides an in-depth definition of an APT \cite{nist2011}. For the purpose of this paper, we consider APT groups as state and state-sponsored threats. For instance, the Daqu, Flame and Gauss are examples of malware used by allegedly state funded APT groups as part of espionage campaigns \cite{Bencsathetal2012}. Additionally, these campaigns, alongside Stuxnet and Red October, display the difficulty of detecting state and state-sponsored APT threats \cite{VirvilisGritzalis2013}.
At the moment, there exists only sparse information on APT groups, and the data remains unstructured and difficult to automatically analyze. \citet{Lemayetal2018} created a survey which contains several pieces of information on APT groups retrieved from public sources, such as the various aliases used for group names and the alleged campaigns conducted. 

\begin{table*}[!ht]
	\centering
	\begin{threeparttable}
		\caption{Revised list of the top 10 APT groups. We gathered information from AT\&T Cybersecurity \cite{attcybersecurity}, MITRE \cite{mitre} and CCN-CERT \cite{ccn-cert} to create this list. The table also reports the alleged group location and the number of unique and shared tools linked to each group.}
		\label{tbl:top_apt}
		\begin{tabular}{L{0.5cm}L{0.5cm}p{2.15cm}L{1cm}L{5cm}L{1.5cm}L{1.25cm}L{1.25cm}}
			\toprule
			\rowcolor{white} & & & & & & & \\
			\rowcolor{white} \multicolumn{2}{c}{\multirow{-2}{0.5cm}{Rank}} & & & & & & \\
			\rowcolor{white} 2020 & 2018 & \multirow{-3}{2cm}{Group Name} & \multirow{-3}{1.25cm}{Number of Aliases} & \multirow{-2}{3.5cm}{Aliases} & \multirow{-3}{1.5cm}{Suspected Location} & \multirow{-3}{1.5 cm}{Number of Unique Tools} & \multirow{-3}{1.5cm}{Number of Shared Tools} \\
			\midrule		
			\rowcolor{Gray} 1 & 1 & \multirow{1}{2.25cm}{Lazarus Group} & 4 & \multicolumn{1}{m{5cm}}{HIDDEN COBRA, Guardians of Peace, ZINC, NICKEL ACADEMY} & DPRK & 16 & 2 \\ 
			\rowcolor{white} & & & & & & & \\
			\rowcolor{white} \multicolumn{1}{L{0.5cm}}{\multirow{-2}{0.2cm}{2}} & \multicolumn{1}{L{0.5cm}}{\multirow{-2}{0.2cm}{$\star$}} & \multicolumn{1}{L{2.15cm}}{\multirow{-2}{2.15cm}{Gamaredon Group}} & \multicolumn{1}{L{1cm}}{\multirow{-2}{0.2cm}{0}} & \multicolumn{1}{m{5cm}}{\multirow{-2}{3cm}{N/A}} & \multicolumn{1}{L{1.1cm}}{\multirow{-2}{1.5cm}{N/A}} & \multicolumn{1}{L{1.25cm}}{\multirow{-2}{0.2cm}{1}} & \multicolumn{1}{L{1.25cm}}{\multirow{-2}{0.2cm}{0}} \\
	
			\rowcolor{Gray} 3 & 7 & Kimsuky & 1 & \multicolumn{1}{m{5cm}}{Velvet Chollima} & DPRK & 0 & 0 \\	
			\rowcolor{white} 4 & 3 & MuddyWater & 2 & \multicolumn{1}{m{5cm}}{TEMP.Zagros, Seedworm} & Iran & 2 & 6 \\ 
			\rowcolor{Gray} 5 & $\star$ & TA505 & 1 & \multicolumn{1}{m{5cm}}{Hive0065} & N/A & 5 & 3 \\				
			\rowcolor{white} 6 & 2 & Sofacy & 11 & \multicolumn{1}{m{5cm}}{SNAKEMACKEREL, APT 28, Sednit, Pawn Storm, Group 74, Tsar Team, Fancy Bear, Strontium, Swallowtail, SIG40, Threat Group-4127 } & Russia & 20 & 4 \\			
			\rowcolor{Gray} 7 & $\star$ & PROMETHIUM & 1 & \multicolumn{1}{m{5cm}}{StrongPity} & N/A & 2 & 0 \\
			\rowcolor{white} 8 & 10 & Turla & 5 & \multicolumn{1}{m{5cm}}{Snake, Venomous Bear, Waterbug, WhiteBear, Krypton} & Russia & 10 & 10 \\	
			\rowcolor{Gray} 9 & 4 & Oil Rig & 3 & \multicolumn{1}{m{5cm}}{IRN2, HELIX KITTEN, APT 34} & Iran & 9 & 11\\	
			\rowcolor{white} 10 & $\star$ & \multirow{1}{2.25cm}{Emissary Panda} & 6 & \multicolumn{1}{m{5cm}}{TG-3390, BRONZE UNION, Threat Group-3390, APT27, Iron Tiger, LuckyMouse} & China & 3 & 12\\						
			
			\bottomrule	
		\end{tabular}
	\end{threeparttable}
	\begin{tablenotes}
		\footnotesize
		\item $\star$ Not in the 2018 top 10 APT groups.
	\end{tablenotes}	
\end{table*}

In addition, there exists a publicly available spreadsheet containing APT groups and their aliases \cite{aptspreadsheet}. Various cyber-experts from several reputable cyber-threat intelligence sources, such as FireEye, CrowdStrike and MITRE \cite{mitre}, regularly contribute to the spreadsheet and it quickly gained popularity amongst the research community for the \emph{ground truth}. There also exist a few open-source sharing methods such as STIX \cite{stix} and TAXII \cite{taxii} to help researchers, but most threat intelligence options require payment \cite{Bouwmaneta2020}.

To further highlight the issues surrounding APT groups, we gathered information from MITRE \cite{mitre}, AT\&T Cybersecurity \cite{attcybersecurity} and CCN-CERT \cite{ccn-cert} to create a list, \Cref{tbl:top_apt}, of the top ten APT groups along with the alleged group location and tools linked to each group. From the table, we see the vast number of aliases and the lack of samples linked to each APT group. For example, there currently exists no known malware linked to the group \textit{Kimsuky}. We also observe the majority of the APTs use both unique malware and open source/shared tools \textit{e.g.}, \textit{Turla} and \textit{Oil Rig} use PsExec but allegedly, different Nation States sponsor them. Therefore, MAA becomes increasingly harder if all groups use identical tools. Furthermore, we remark there exist no APT groups on the list allegedly sponsored by a Western or Five Eye nation\footnote{The Five Eyes consist of United States, United Kingdom, Canada, Australia and New Zealand}. We believe the source of the data, predominantly American Threat Intelligence companies, might introduce some bias to the list as their focus aligns to the threat actors of Western or Five Eye Nation. However, threat actors target and belong to a variety of countries. Finally, we remark from 2018 to 2020 six APT groups remain in the top 10. This shows the longevity of the groups despite an increase in public attribution.

\subsection{Binary Similarity and YARA Rules}
Currently, malware analysts use YARA rules\footnote{YARA is a pattern matching tool with a rule based syntax which allows the discovery of specific signatures \cite{yara}.} for recognizing and attributing malware samples. YARA rules tend to identify shellcode and code reuse for linking samples and not authorship style which is akin to the binary similarity problem, \textit{i.e.} comparing how much shared code exists between binaries \cite{UlhaqCaballero2019} or searching binaries for code cloning \cite{Farhadietal2015}. Using similarity for attribution is not foolproof and in many cases can be lead to false accusations \cite{BartholomewGuerrero-Saade2016}. It usually also requires analyzing all of the binary whereas identifying author style can be performed on smaller code fragments. 

Even though an analyst must write a rule based on their research of each unique sample (meaning YARA rules remain as labor-intensive to most manual malware analysis methods), they provide a much easier and quicker solution to the current MAA systems. Research by \citet{BassatCohen2019} shows the ease of using YARA rules in the ``wild'' for clustering malware similarities between alleged Russian APTs. However, the same research also shows YARA rules rely on unpacked samples to trigger the identified traits within the YARA rules and this is similar to current MAA systems. More recently, \citet{Raffetal2020} tackle the labor-intensive problem and develop the state-of-the-art to automatically generate YARA rules using malware.  Similar to the research by \citet{BassatCohen2019}, Kaspersky developed a Threat Attribution tool based on APT malware binary similarity \cite{Kaspersky2020}.

\subsection{Binary Authorship Attribution}
MAA is a subset of the binary authorship attribution (BAA) problem. BAA applies to other tasks, such as plagiarism and intellectual property rights. In these cases, we know all the authors beforehand, \textit{e.g.,} the students in a programming class. In contrast, malware authors wish to remain undisclosed due to the illegality and secrecy of the underground market within which they operate \cite{Afrozetal2014}. When we know all the possible authors, we call this \textit{Closed World (Assumption)}, otherwise \textit{Open World (Assumption)} \cite{MoorePham2015}. All of the authorship systems reviewed in this work use \textit{Closed World Assumption (CWA)}. From a data modeling perspective, this prevents understanding the real world context of authorship attribution. However, the CWA mitigates some of the challenges such as quantifying some of the unknowns, \textit{i.e.,} the total number of authors. Mitigating some of challenges can help with exploring other authorship objectives.

There exists varying objectives of authorship attribution set out by \citet{Kalgutkaretal2019}. These consist of \textit{identification} - linking a binary to an author, \textit{clustering} - grouping stylistic similarities, \textit{evolution} - tracking stylistic changes over time, \textit{profiling} - understanding stylistic characteristics, and \textit{verification} - checking for adversarial tampering.
We focus on identification as the other objectives can be a by-product of the research on authorship \textit{identification} and \textit{identification} forms the basis of understanding authorship style which the other objectives rely on.

Within all authorship attribution objectives, we must consider if the goal is \textit{Single} or \textit{Multiple} authorship. The single authorship attribution problem assumes only one author for every piece of binary. Conversely, the multiple authorship attribution problem assumes multiple authors created the binary.  Assuming single authorship of binaries which multiple authors created is likely to make any attribution system incorrectly learn authorship style and lead to potential attacks on the system which we explore in the next section.

\subsection{Adversary Techniques} \label{sec:adversary}
Most BAA work assumes the authors are unaware of an attribution system being in place. Few works consider authors using adversarial techniques to influence the output of the attribution system. This requires attackers to first identify which features appear easier to manipulate to affect the output of the attribution system. Some of these attacks are aimed at the learning phase (\textit{e.g,}\emph{ training set poisoning}). However, most existing binary modification attacks are aimed at evading the attribution system at run-time (\emph{evasive attacks}). \citet{Mengetal2018} describe three evasive attacks namely; (i) the confidence-loss attack, (ii) the untargeted attack and (iii) the targeted attack. The confidence-loss attack defeats an attribution system by removing any traces of author style to ensure it predicts no author label for the binary. The untargeted attack attempts to make the prediction of the attribution system as any other author than itself. The targeted attack tries to convince the attribution system the binary belongs to a pre-chosen author other than the attacker. 
We deem the confidence-loss attack as unsophisticated as most malware authors try this by default to remain anonymous and maintain their privacy.  Whereas we class the other two attacks as sophisticated and we believe APT groups are more likely to implement these attacks. 

\subsubsection{Unsophisticated attacks.} We deem these attacks to be obfuscation techniques authors use to hide their identity and fool malware detection systems. Common obfuscation techniques include \emph{encryption} and \emph{packing}. The use of encryption prevents easy analysis. The adversary encrypts the main function to prevent static analysis on the malware. The program initially calls a function to decrypt itself upon runtime. This function requires a decryption key which the author either stores at a remote location (such as a Communication and Control server) or hides in the malware delivery method (such as a phishing email). Otherwise, storing the key in the malware file allows for the malware analysts to decrypt it.

Malware authors use \emph{packing} to evade analysis and detection systems. The developer compresses the binary to hide the functionality of the binary. A packed binary contains a small amount of code which enables the binary to decompress itself at runtime. A packed version of a binary appears as a completely different version to the original binary, which allows adversaries to trick defense systems such as anti-virus software. The majority of packed binaries require manually unpacking before applying static analysis techniques. However, there exist automatic tools such as Un\{i\}packer which unpacks common packing tools such as UPX, ASPack, PEtite and FSG \cite{unipacker}. Authorship attribution systems either require the samples unpacked to extract author style or they apply their process to packed binaries to test if authorship style remains after packing.

\subsubsection{Sophisticated attacks.} \emph{False flags} used by APTs to imitate other groups \cite{BartholomewGuerrero-Saade2016} are the primary example of sophisticated attacks currently in use. \citet{Simkoetal2018} considered the idea of \emph{imitating programmer style} for source code authorship attribution. This led to the definitions of \textit{Forgery} and \textit{Masking} techniques. The \textit{Forgery} technique describes the process an adversary employs to create a program which the attribution system outputs as a different APT group. For example, we describe a targeted attack by A on B (involving an innocent party C) when A successfully convinces B that C performed the attack. If A convinces B any other attacker executed the attack, then we class the attack as untargeted.
\textit{Masking} is when an adversary manages to hide as the original author of a program it has modified. For example, an attacker wants to add malicious code into an open source project without the original authors knowing. Similarly to Forgery, masking can either be targeted or untargeted.
\citet{Matyukhinaetal2019} develop such an attack to five state-of-the-art source code authorship attribution models by learning authorship style from data collected from open source repositories. They create three types of source code transformation attacks based on capturing author style to create both targeted and untargeted attacks. Similarly, \citet{Quiringetal2019} construct a Monte-Carlo Tree search to transform source code for both targeted and untargeted attacks on two state-of-the-art source code authorship attribution systems. Interestingly they both circumvent the authorship attribution system by \cite{Caliskanetal2015} using different approaches.
These attacks on source code authorship attribution systems show MAA systems are likely to face similar attacks and so any system must consider such attacks.

\section{Malware Binary Authorship Attribution} \label{sec:systemization}

We reviewed papers in the subject field over the last decade to identify relevant systems and research applicable to MAA. Our search criteria looked for work which addressed the problem of binary and malware authorship attribution. We omitted any papers which performed a binary classification on malware and contained no significant contribution on authorship styles to malicious files, \textit{e.g.,} we omit the paper \cite{Laurenzaetal2017} as this classifies malware into APT group or non-APT group but we include \cite{Laurenzaetal2018} as this classifies malware into specific APT groups. We identified eighteen papers which possess a significant relationship with MAA, and we contacted all authors whose systems were not publicly available. We received a mixture of responses. Some systems had contractual obligations to prevent them from being shared, others did not wish to share their system or said their system shall be made available in the future. 
On top of the eighteen papers, we identified the survey by \citet{Alrabaeeetal2017} which evaluates the systems in \cite{Rosenblumetal2011, Alrabaeeetal2014, Caliskanetal2018}. Although this paper provides no new system it helps provide added insight on the systems they evaluated in the context of malware. We focus on: (i) \emph{data modeling techniques}, (ii) \textit{datasets}, and (iii) \textit{features}. We decided on these three areas as they represent the key components in building analytical systems for understanding large data.

In Section \ref{sec:data_modeling}, we first classify the \emph{data modeling techniques} used in these works into five categories: (i) \textit{classification} techniques identify whether a piece of malware belongs to known set of groups; (ii) \textit{clustering} techniques enable us to group malware into authors based on underlying data trends; (iii) \textit{anomaly detection} methods allow us to label malware based on malware not conforming to a known group or category; (iv) \textit{structured prediction} methods predict structured objects for example within a binary file we can identify a structure for an author based on assembly language; (v) \textit{non-machine learning methods} include alternative probabilistic or manual methods.	

In Section \ref{sec:datasets}, we categorize the works based on the \textit{datasets} used within the systems. Specifically, we divide the  \textit{datasets} into benign source code, benign binaries and malware binaries to match the current approach by researchers. The benign software approach uses compiled source code from known authors and the malware approach uses predominantly APT malware. In Section \ref{sec:features}, we explore malware author style and derive a categorization of author features which we use to compare the eighteen BAA systems.

\begin{table}[!t]
	\rowcolors{1}{white}{Gray}
	\centering
	\caption{A list of known Data Modeling Techniques used to tackle the binary authorship problem published between 2011 and 2019. There exist five categorizes of techniques: (i) Classification; (ii) Clustering; (iii) Anomaly detection; (iv) Structured prediction; and (v) Non-machine learning methods.}
	\label{tbl_ml_algorithms}
    \begin{tabular}{p{3cm}p{6.75cm}p{5.5cm}}
		\toprule
		\multirow{2}{3cm}{Data Modeling} & \multirow{2}{6.75cm}{Algorithm} &  \multicolumn{1}{p{1cm}}{\multirow{2}{3.5cm}{Attribution System}} \\ 
		\addlinespace[1.5em]
		\midrule
		\cellcolor{white} & Deep/Artificial Neural Networks (DNN/ANN) & \cite{Rosenbergetal2017}, \cite{Rosenbergetal2018}, \cite{MengMiller2018}, \cite{Alrabaeeetal2019a}, \cite{Alrabaeeetal2019b}\\
		& Tree Bagging (TB) & \cite{Hongetal2018} \\
		
		\cellcolor{white} & Random Forests (RF) & \cite{Hendrikse2017}, \cite{Caliskanetal2018}, \cite{Hongetal2018}, \cite{Gonzalezetal2018}  \\
		
		& Support Vector Machine (SVM) & \cite{Rosenblumetal2011},  \cite{Meng2016}, \cite{Mengetal2017}, \cite{Caliskanetal2018}, \cite{Hongetal2018}, \cite{Kalgutkaretal2018}, \cite{MengMiller2018}\\
		
		\cellcolor{white} & Bayesian Classifiers (\textit{e.g.}, Na{\"i}ve Bayes (NB)) &  \cite{Hendrikse2017}, \cite{Hongetal2018} \\
		\multirow{-7}{2cm}{Classification} & Large Margin Nearest Neighbor (LMNN) & \cite{Rosenblumetal2011} \\
		\midrule

		\cellcolor{white} & K-Means Clustering & \cite{Rosenblumetal2011}, \cite{Alrabaeeetal2019a}, \cite{Alrabaeeetal2019b} \\ 
		\cellcolor{white} \multirow{-2}{*}{Clustering} & Multi-View Fuzzy Clustering & \cite{Haddadpajouhetal2020} \\ \midrule				
				
		Anomaly Detection & Isolated Forests (IF) & \cite{Laurenzaetal2018} \\
		
		\midrule			
		\cellcolor{white} Structured Prediction & Conditional Random Fields (CRFs) & \cite{Mengetal2017}, \cite{MengMiller2018} \\
		\midrule

		& Dissimilarity Algorithm & \cite{Alrabaeeetal2014} \\
		\cellcolor{white} & Manual Analysis & \cite{Marquis-Boire2015} \\

		\multirow{-3}{*}{Non-Machine Learning} & Attribution Weighting & \cite{Alrabaeeetal2018a} \\
		\bottomrule	
	\end{tabular}	
\end{table}	
	
\subsection{Data Modeling Techniques}\label{sec:data_modeling}
We present all the techniques used from the reviewed papers in \Cref{tbl_ml_algorithms}. From the table, we see fifteen of the systems use various Machine Learning (ML) methods. We also notice the majority of ML methods favor the classification problem. We believe the reason for this lies in the easier approach of solving the closed-world problem using labeled source code data which we show in Section \ref{sec:datasets} and Section \ref{sec:keyfindings}. Research on source code authorship attribution mirrors the same pattern \cite{Kalgutkaretal2019}. 

\citet{Hongetal2018} uniquely explore more than two classification algorithms and conclude Random Forest (RF) and Support Vector Machine (SVM) as the most suitable candidates for solving the problem due to their enhanced performance against the other five techniques they tested. This concurs with the rest of the field \cite{Hendrikse2017,Caliskanetal2018,Kalgutkaretal2018,Meng2016,Gonzalezetal2018}.  Seven papers consider three alternative ML methods: clustering, anomaly detection and structured prediction techniques \cite{Rosenblumetal2011,Laurenzaetal2018,Mengetal2017,MengMiller2018,Alrabaeeetal2019a, Alrabaeeetal2019b,Haddadpajouhetal2020}. We explore these further as they show promise towards the open-world problem. 

In detail, \citet{Rosenblumetal2011} use a SVM classifier within their single-author closed-world model and they extend this solution to the open-world problem by using a k-mean clustering technique to cluster binaries based on previously built author profiles. For this, they change their original classifier to the Large Margin Nearest Neighbor (LMNN) as this aids building author profiles. \citet{Laurenzaetal2018} approach APT triaging by identifying outliers of APT style within malware using Isolated Forests (IF). 

\citet{Mengetal2017,MengMiller2018} extend the multiple author feature discovery work (\cite{Meng2016}) by using Conditional Random Fields (CRFs) applied to the assumption multiple authors code consecutive basic blocks. In this scenario, CRFs outperform SVMs. Continuing this assumption, \Citet{MengMiller2018} explore the use of Deep Neural Networks directly on the binaries' raw bytes without any analysis or feature extraction process. \citet{Rosenbergetal2017,Rosenbergetal2018} also consider the use of Artificial Neural Networks for classifying binaries to authors. \citet{Alrabaeeetal2019a,Alrabaeeetal2019b} use convolutional neural networks to cluster author style and then use a classifier to determine if a piece of malware belongs to an author cluster. Finally, \citet{Haddadpajouhetal2020} choose a multi-view fuzzy clustering model to group malware into APT groups based on identifying loosely defined patterns among binary artifacts.

Alternative non-ML methods used to solve the BAA problem also use features which identify author style. Both \citet{Alrabaeeetal2014} and \citet{Alrabaeeetal2018a} use probabilistic methods such as dissimilarity algorithms and a novel attribution weighting formula respectively. \citet{Marquis-Boire2015} propose a pipeline driven from manual malware analysis.

\subsection{Current Datasets}\label{sec:datasets}
Datasets remain a key part of any analysis process due to the necessity of identifying binary specific trends within the data. We summarize the current sources used within the eighteen systems reviewed. We split the dataset analysis into two sections: (i) Benign Source Code and Binaries; and (ii) Malware Binaries. Afterwards, we provide an overall comparison of the datasets.

\subsubsection{Benign Source Code and Binaries}\label{sec:benigndatasets}
Due to the lack of author labeled binaries, the majority of the research in BAA uses source code from student competitions and then compiles it using a variety of compilers to create a ground truth binary dataset. This approach allows researchers more control on the \textit{cleanliness} of the dataset. Specifically, this provides researchers with greater certainty on the verification of the ground truth. In addition, this provides the ability to choose which complexities the toolchain process introduces, artificially create larger datasets by using multiple toolchain processes and link author styles learned from source code stylometry. Consequently, this approach leads to datasets which fail to represent the real world. They tend to remain static and not evolve alongside author styles. Additionally, these datasets add extra time to consider all the different toolchain combinations to account for the various compilation methods. Researchers also choose the datasets to consist of only C and C++ languages due to the popularity of the programming languages \cite{tiobe}. However, malware generally consists of various languages. We describe the four main sources below.

\paragraph*{Google Code Jam (GCJ) \cite{googlecodejam}} Since 2008, this worldwide student competition runs annually and the organizers publish all the problems and solutions for anyone to download. There exist multiple benefits for using the GCJ dataset for authorship identification. Firstly, all the participants code similar programs and this allows researchers to focus purely on author style and not program functionality. Secondly, the dataset consists of diverse authors from all over the world.
Thirdly, GCJ offers substantial prizes to the participants meaning they must know their identity. Hence, there is no necessity for the participants to hide their author style unlike malware creators. In general, the overall quality of the submissions varies as not all the samples compile meaning researchers must clean the dataset before using it.

\citet{Hendrikse2017} uses the script written by \citet{Caliskanetal2018} to obtain the GCJ dataset. However, they both use different subsets of the same dataset for testing and training their attribution systems.
\citet{Alrabaeeetal2019b} use the GCJ dataset to build synthetic binaries from multiple authors by combining the source codes of the various entries. They construct binaries consisting of between two and eight authors. However, this method introduces the issue of distinct separation between the various author styles within the binaries. Therefore, we believe this method constructs a poor dataset for training BAA systems due to the cleanliness allowing the systems to easily distinguish between the authors. However, the dataset provides an opportunity to test systems and evaluate whether they actually perform highly on such a clean dataset.

\paragraph*{GitHub \cite{github}} This is a hosting site for software development which uses \textit{git}, an open source version control platform. GitHub encourages agile development for software projects and allows multiple authors to edit and contribute to various repositories whilst recording the contribution of each user. %
\citet{Mengetal2013} created the tool \textit{git-author} to tackle the attribution of GitHub repositories to each author. This enabled them to create a labeled dataset for multiple authorship attribution. Three works use git-author for the ground truth of their attribution system \cite{Meng2016, Mengetal2017,MengMiller2018}. Additionally, the GitHub community ranks each repository out of five stars which \citet{Caliskanetal2018} use to judge programmer ability. In this work, they build their GitHub dataset using only repositories containing at least two hundred lines of code and they omit any forked repositories or any named ``Linux'', ``kernel'', ``OSX'', ``LLVM'' or ``next''. They state this ensures a sufficient amount of code exists to learn author style and it also reduces the amount of shared code within their GitHub dataset. \citet{Alrabaeeetal2019b} collect fifty C/C++ projects where between 50 and 1,500 authors contributed to each project. Introducing a high number of authors potentially saturates author style boundaries as there exists some natural cross-over with author style making it even harder to distinguish between the distinct authors.

Plenty of disadvantages exist from using this data source for malware attribution. Firstly, the majority of repositories are benign projects and malware authors are unlikely to use popular open source repositories for malware development. Secondly, the openness of GitHub allows anyone to clone the code and in turn author style. Finally, it opens up the code to the potential attack where an adversary modifies the code without the repository owner noticing through author style imitation \cite{Simkoetal2018}.

\paragraph*{Planet Source Code \cite{planetsourcecode}} This platform hosts source code and claims to host 4.5 million lines of code and this includes approximately 200,000 lines of C/C++ code. When a user uploads their code to the site, they rank their own skill level choosing the option of unranked, beginner, intermediate or advanced. Other site members then rank each submission for the various awards the site offers. The combination of both these ranking methods provides site users with confidence in the coding standard. Similar to previous data sources, there exists the assumption any uploaded source code belongs to the user who uploads the code.

\paragraph*{Other Benign Sources.} In addition to the three public repositories above, \citet{Rosenblumetal2011,Alrabaeeetal2018a} use student coursework. \citet{Alrabaeeetal2018a} assume the source code author refers to the student who submitted the coursework, whereas the dataset used by \citet{Rosenblumetal2011} included submissions where the students worked in pairs. To mitigate this issue, they performed manual analysis to identify a single author for each program. Alternatively, academics use plagiarism detectors on coursework submissions to identify where students cheated and this provides a form of ``authorship attribution''. However, plagiarism checkers fail to check for contributions from unknown third parties \cite{Albluwi2020, Foltyneketal2020}. In comparison, for MAA we must consider methods to identify unknown programmers/malware authors due to the ``underground'' behavior exhibited \cite{Afrozetal2014}. \citet{Rosenblumetal2011} state the students in their dataset received skeleton code which potentially influenced the students' programming style even though they attempted to remove all the skeleton code from the samples. In comparison to malware authors, the students must identify themselves to receive a score for their coursework and therefore are likely to refrain from implementing methods to hide their author style. Unfortunately, data protection policies prevent both \citet{Rosenblumetal2011,Alrabaeeetal2018a} from sharing the datasets.

\citet{Kalgutkaretal2018, Gonzalezetal2018} created a benign Android application dataset using applications from stores such as Google Play Store, Appland, Anzhi, Aptoide, Fdroid, MoboMarket, Nduoa, Tincent and Xiaomi for which they attribute by using the private certificates from the signed APK files. Additionally, \citet{Gonzalezetal2018} use the store called 3gyu. As well as the previous application stores, both these papers use APK files from GitHub and an on-line collaborative system called Koodous.

\paragraph*{Toolchain} \label{sec:toolchain} The common toolchain approach uses multiple compilers and optimization levels. However, every combination of compiler and optimization level used produces a unique binary sample from the same source code. This generic approach excludes the use of varying obfuscation tools and modifications which create further unique binaries. There exist six papers \cite{Hendrikse2017, MengMiller2018,Alrabaeeetal2017,Alrabaeeetal2018a,Alrabaeeetal2019a,Alrabaeeetal2019b} which create a dataset using multiple compilers from both open source and commercial sources, such as Clang \cite{clang}, GNU \cite{gnu}, ICC \cite{icc}, LLVM \cite{llvm}, Microsoft Visual Studio \cite{visualstudio} and Xcode \cite{xcode}. A sophisticated malware developer might create a customized compiler yet this remains unlikely due to the deterrence of the complexities of compiler design and it is a unique identifier.
The optimization functionality of compilers decreases the program's runtime, but at the same time it increases the compilation duration. Programmers consider a cost-benefit analysis when deciding which level of optimization to perform. Similar to using different compilers, using varying optimization levels affects author style. Eight papers consider at least one optimization level within their research to account for the effect of optimization on author style \cite{Caliskanetal2018,Meng2016,Mengetal2017,Hendrikse2017,MengMiller2018,Alrabaeeetal2017,Alrabaeeetal2018a,Alrabaeeetal2019b}. However, there still requires further understanding of the impact of toolchains on author style.

\subsubsection{Malware Binaries}\label{sec:malwaredatasets}
Creating an author labeled malware dataset echoes similar difficulties in creating a malware family labeled dataset \cite{Sebastianetal2016}. We show particular interest in APT malware as APT groups tend to use sophisticated adversarial techniques. To the best of our knowledge, there exist two attempts to create a large APT labeled dataset. \citet{Laurenzaetal2018} create a list of APT groups and use these to scan publicly available reports written by threat intelligence companies, government departments, anti-virus and security companies for related malware hashes. They use these hashes to download the samples from sources such as VirusTotal. They store this dataset on GitHub \cite{aptdataset}. For the purpose of their paper \citet{Laurenzaetal2018} use a subset of \cite{aptdataset} consisting of 19 APT groups and over 2000 malware samples. Due to the unavailability of the exact dataset used in \citet{Laurenzaetal2018}, we analyzed the GitHub dataset \cite{aptdataset}. The second attempt to create an APT malware dataset is by \textit{ ``cyber-research''} which they store on GitHub \cite{aptdataset2}. The dataset contains 3594 malware samples\footnote{\textit{ ``cyber-research''} also include information on a further 855 samples which they could not obtain.} which are related to twelve APT groups and are allegedly sponsored by five different nation-states. Similarly to \citet{Laurenzaetal2018}, \textit{ ``cyber-research''} collect the malware samples using open source threat intelligence reports from multiple vendors and then downloaded from VirusTotal. However, \textit{``cyber-research''} omit the method they used to label the malware hashes from the 29 sources and so researchers have no assurances on the validity of the label. We note \citet{Haddadpajouhetal2020} use a subset of \cite{aptdataset2}, focusing on five groups namely APT1, APT3, APT28, APT33, and APT37. In both cases, we observed general issues with creating labeled APT malware datasets:

\begin{itemize}
	\item\textbf{APT group names used for a single APT group often differ which leads to multiple aliases and not knowing which common name to use as the label.} In some cases, different groups share the same aliases. Either researchers linked multiple APT groups to the same nation or multiple APT groups potentially collaborated together. This makes it difficult to create a single list which contains a one-to-one relationship between sample and group. This problem relates to the one solved by \citet{Hurieretal2017}, who produce a distinct naming dataset for malware family names as anti-virus vendors use their own naming conventions.
	\item\textbf{Reports on APT groups often reference multiple groups when the researchers compare or link groups.} Therefore, researchers must take extra care when automatically extracting labels from the reports. For example, within \cite{aptdataset} there exist the same reports linked to differing APT groups.
\end{itemize}

Due to these problems and the availability of APT datasets, some authors use alternative malware datasets or obtain datasets from private sources. \citet{Alrabaeeetal2017} obtain malware from their own Security Lab (Zeus and Citadel malware), from Contagio (Flame and Stuxnet malware) and from VirusSign (Bunny and Babar malware). They omit the method they use to determine the ground truth for this dataset. It appears \citet{Alrabaeeetal2018a} use the same dataset and they state they manually determined the labeling. \citet{Alrabaeeetal2019a} use a similar dataset but they add samples of the Mirai botnet to the dataset.
The Microsoft Malware Classification Challenge dataset \cite{Ronenetal2018} provides an alternative popular malware source. Three works use subsets of this dataset \cite{Alrabaeeetal2014,Alrabaeeetal2019a,Alrabaeeetal2019b}. The dataset by \citet{Ronenetal2018} contains nine malware families\footnote{Ramnit, Lollipop, Kelihos\_ver3, Vundo, Simda, Tracur, Kelihos\_ver1, Obfuscator.ACY, Gatak.} and currently there exist no links between the nine malware families and APT groups \cite{mitre}.

Four papers omit their malware sources and they all use cyber security experts to label their datasets \cite{Marquis-Boire2015, Rosenbergetal2017, Hongetal2018, Rosenbergetal2018}. Only the works by \citet{Rosenbergetal2017,Rosenbergetal2018} use datasets with labels representing the nation states which the APT groups are allegedly from or backed by. In particular, they use malware allegedly from or backed by two countries, namely Russia and China. \citet{Rosenbergetal2018} state the dataset consists of four unique malware families in the training set\footnote{Net-Traveler and Winnti/PlugX both allegedly China and Cosmic Duke and Sofacy/APT28 both allegedly Russia.}, with 400 samples from each family, and they use two unique malware families in the testing set\footnote{Derusbi allegedly China and Havex allegedly Russia.}, with 500 samples from each family. \citet{Marquis-Boire2015} use the smallest dataset containing only three samples (NBOT, Bunny and Babar) which they claim belong to the APT group named Snowglobe\footnote{This group allegedly associates with France.}.
The alternative approaches to MAA by \citet{Kalgutkaretal2018} and \citet{Gonzalezetal2018} look to explore Android malware datasets. These works offer an interesting approach towards labeling the malware by the private certificates from the signed APK files. This approach is unique to Android malware and therefore fails to generalize. \citet{Gonzalezetal2018} also perform manual analysis as they consider a lot more malware including APK files from Virus Total, Hacking Team and the Drebin dataset.

\subsubsection{Comparison of Datasets} \label{sec:datasetcomparison}
\Cref{tbl:datasets} provides an overview of all the different datasets used within the current research. We organized \Cref{tbl:datasets} as follows: we clustered all the columns relating to \emph{benign source code and binaries} and then incorporate our discussion on \emph{malware binaries} under the same titled column; we kept the \emph{Ground Truth} column separate to highlight the various methods across both benign and malware datasets; finally, we recorded the largest number of authors and binaries considered in each work. We note the work by \citet{Alrabaeeetal2019b} appears twice in the table due to the work using two distinct datasets for tackling the single and multiple author problem.

Overall, we observe the lack of systems using malware as the sole dataset. Among those papers which use malware, researchers use binaries collected from various sources and samples. This variety means there exists little overlap between the different datasets preventing true system comparison. In most cases, few samples exist for each author which makes it extremely hard for an attribution system to pick up on author style trends.

Limited datasets exist for the multiple authorship problem. Currently, researchers use benign source code from GitHub repositories to compile multiple author binaries or they synthetically create them from single author benign source code. Both these methods create binaries which represent the extremes of author style within a binary: the GitHub binaries contain many author styles distributed across the binary \cite{Dauberetal2019} and the synthetic binaries contain multiple author style separated into distinct sections within the binary \cite{Alrabaeeetal2019b}. Additionally, both datasets lack specific malware author style traits.

\begin{landscape}
	\begin{table*}
		\rowcolors{1}{Gray}{white}
		\centering
		\begin{threeparttable}
			\caption{A summary of the largest datasets and sources used within the papers we reviewed published between 2011 and 2019. We include the toolchain process for the datasets created from source code and the method of author labeling to determine the ``Ground Truth''.} 
			\label{tbl:datasets}
			\begin{tabular}{p{3.5cm}rL{1cm}L{1cm}L{0.75cm}L{0.75cm}p{1.5cm}L{1.85cm}p{1.55cm}L{1.1cm}p{1cm}rr}
				\toprule
				\rowcolor{white}
				& & \multicolumn{7}{m{7cm}}{Benign Source Code and Binaries} &  &  &  & \\
				\multirow{-2}{*}{Paper} & \multirow{-2}{*}{Year} & GCJ & GitHub & Planet & Other\tnote{a}  & Languages & Optimization\tnote{b} & Compilers\tnote{c} & \multirow{-2}{1cm}{Malware Binaries} & \multirow{-2}{1.1cm}{Ground Truth\tnote{d}} & \multirow{-2}{*}{Authors\tnote{e}} & \multirow{-2}{*}{Binaries\tnote{e}}  \\
				\midrule
				\citet{Rosenblumetal2011} & \citeyear{Rosenblumetal2011} & \cmark & & & \cmark & C/C++ & & G & & $\star$ & 191 & 1,747 \\
				\citet{Alrabaeeetal2014} & \citeyear{Alrabaeeetal2014} & \cmark & & & & C/C++ & & & & $\star$ & 7 & $\square$ \\ 
				
				\citet{Marquis-Boire2015} & \citeyear{Marquis-Boire2015} & & & & & & & & \cmark & $\bullet$ & 3 & 3\\

				\citet{Meng2016} & \citeyear{Meng2016} & & \cmark & & & C/C++ & 1 & G &  & $\triangleleft$ & 282 & 170 \\
				
				\citet{Mengetal2017} & \citeyear{Mengetal2017} & & \cmark & & & C/C++ & 1 & G &  & $\triangleleft$ & 284 & 169 \\
				
				\citet{Rosenbergetal2017} & \citeyear{Rosenbergetal2017} & & & & & & & & \cmark & $\bullet$ & 2 & 4,200\\
				
				\citet{Hendrikse2017} & \citeyear{Hendrikse2017} & \cmark & & & & C/C++ & 2 & GLM & & $\star$ & 14 & 1,863 \\
				
				\citet{Alrabaeeetal2017} & \citeyear{Alrabaeeetal2017}  & \cmark & \cmark & & & C/C++ & 1 & GIM\tnote{f} X & \cmark & $\star\diamond\triangleright$ & 1,000 & $\square$ \\
				
				\citet{Caliskanetal2018} & \citeyear{Caliskanetal2018} & \cmark & \cmark & & & C/C++ & 3 & G & & $\star$ & 600 & 5,400 \\
				
				\citet{MengMiller2018} & \citeyear{MengMiller2018}  & & \cmark & & & C/C++ & 5 & GIM &  & $\triangleleft$ & 700 & 1,965 \\
				
				\citet{Hongetal2018} & \citeyear{Hongetal2018} & & & & & & & & \cmark & $\bullet$ & 7 & 1,088\\
				
				\citet{Alrabaeeetal2018a} & \citeyear{Alrabaeeetal2018a} & \cmark & \cmark & \cmark & \cmark & C/C++ & 2 & GICM & \cmark & $\star\diamond$ & 23,000 & 103,800 \\
				
				\citet{Rosenbergetal2018} & \citeyear{Rosenbergetal2018}  & & & & \cmark & & & & \cmark & $\bullet$ & 2 & 4,200\\
				
				\citet{Kalgutkaretal2018} & \citeyear{Kalgutkaretal2018} & & \cmark & & \cmark & Java\tnote{g} & & & \cmark & $\diamond\bullet$ & 40 & 1,559\\
				
				\citet{Gonzalezetal2018} & \citeyear{Gonzalezetal2018} & & \cmark & & \cmark & Java\tnote{g} & & & \cmark & $\diamond\bullet$ & 30 & 420\\
				
				\citet{Laurenzaetal2018} & \citeyear{Laurenzaetal2018} & & & & & & & & \cmark & $\bullet$ & 19 & 2,000+ \\
				
				\citet{Alrabaeeetal2019a} & \citeyear{Alrabaeeetal2019a} & \cmark & \cmark & & & C/C++ & & GICM & \cmark & $\star\bullet$ & 21,050 & 428,460 \\
				
				\citet{Alrabaeeetal2019b}  & \citeyear{Alrabaeeetal2019b} & \cmark & \cmark & & & C/C++ & 2 & GICM & \cmark & $\star\bullet$ & 1,900 & 31,500 \\
				
				\citet{Alrabaeeetal2019b}  & \citeyear{Alrabaeeetal2019b} & \cmark & \cmark & & & C/C++ & 4 & GICM & & $\star$ & 350 & 50 \\
				
				\citet{Haddadpajouhetal2020}  & \citeyear{Haddadpajouhetal2020} &  & & & & & & & \cmark & $\bullet$ & 5 & 1200 \\				
				
				\bottomrule 
			\end{tabular}
			\begin{tablenotes}
				\footnotesize
				\item[a] Other sources for benign datasets where \cmark[0] means they state the source
				\item[b] Number of Optimization Levels used (blank means paper does not state/consider)
				\item[c] Compilers used: G - GCC/g++ I - ICC L - LLVM C- Clang M - Microsoft Visual Studio X - Xcode
				\item[d] Ground Truth Method: $\star$ - Source Code Author $\diamond$ - Manually Determined $\triangleleft$ - \textit{git-author} \cite{Mengetal2013} $\triangleright$ - Undisclosed $\bullet$ - Cyber Security Experts/Malware Analysis Reports
				\item[e] Largest Dataset Stated
				\item[f] \citet{Alrabaeeetal2017} state they use Visual Studio in their methodology but include no dataset details.
				\item[g] Android APK Files
				\item $\square$ - Not Disclosed.
			\end{tablenotes}
		\end{threeparttable}
	\end{table*}
\end{landscape}

In terms of authorship attribution, the researchers treat the APT binaries as single author which importantly introduces false author style links.
The three most promising APT datasets created by \citet{Laurenzaetal2018,aptdataset2} and \citet{Rosenbergetal2018} exhibit flaws. The dataset by \citet{Laurenzaetal2018} contains many APT groups but few samples per group whereas the datasets by \citet{Rosenbergetal2018} contains fewer groups but more samples per group. The dataset by \citet{aptdataset2} lacks assurances surrounding the labeling process.

The issue of verifying the ground truth of the labels of the malware datasets still requires investigating. Source code authors appear easier to distinguish \cite{Kalgutkaretal2019}. In comparison, the majority of malware requires manual analysis or cyber security experts. In the case of malware from the campaign titled `Olympic destroyer', the threat actor used \emph{false flags} to trick analysts into arriving at multiple attribution hypotheses. The original malware authors included specific code reuse from previous campaigns by other attackers. Additionally, they tried to confuse malware analysts by using different spoken language within the comments, user interface and function names. Various analysts discovered the various artifacts throughout the malware at different times and this led to attribution to groups from Russia, Iran, China and North Korea \cite{BartholomewGuerrero-Saade2016}.

Fundamentally, using different datasets means each approach answers slightly different research questions. Furthermore, this suggests a lack of sharing and effort across the research field to try to solve the same problems. In Section \ref{sec:new_dataset}, we hope to change this through the creation of an APT malware dataset which addresses the limitations and shortcomings we identify and publish this for the community to use for future research.

\subsection{Author Features}\label{sec:features}

Capturing author style provides the key to identification. The majority of the state-of-the-art methods determine author style through extracting multiple features and then completing feature ranking experiments using their data modeling techniques (\Cref{tbl_ml_algorithms}) on their chosen ground truth dataset (\Cref{tbl:datasets}). Researchers tend to extract various features based on domain expert knowledge or previous research. In some cases, the papers \cite{Alrabaeeetal2019b,Mengetal2017} use features extracted directly from the binary through either vector or image representation for some experiments. In these cases, it remains unclear which features the model actually uses for author style which presents a gap in the research area. Going forward we review only those specific features which the papers explicitly stated. 

Many of the state-of-the-art BAA systems rely on the area of \emph{code stylometry} research to provide a starting point for features related to author style. Code stylometry features belong to three categories of \textit{lexical}, \textit{syntactic} and \textit{semantic}. However, they omit any features from code execution. To mitigate this, \citet{Kalgutkaretal2019} propose that researchers capture author style from \textit{behavioral and application dependent} characteristics. 

\subsubsection{Malware Author Style} \label{sec:style} Malware authors tend to have unique goals \cite{Reynolds2020} which we can use to help determine the author style and extract features aimed towards capturing the goal of the malware author. \citet{Marquis-Boire2015} remain the only paper to specifically consider malware features for author style through their aim of identifying credible links between APT malware. In particular, they pick up on malware programming style by APT groups such as the use of stealth, evasion and data ex-filtration techniques. \citet{Kaspersky2020} discuss similar themes from their binary similarity research but also widen the search to toolkits, exploits and targeted victim. 
We extend the ideas from these works with our previous discussion to devise five macro-categories, namely \emph{strings}, \emph{implementation}, \emph{infrastructure}, \emph{assembly language} and \emph{decompiler} to compare the state-of-the-art systems in Section \ref{sec:macro_features} along with providing further explanations of each category.

\begin{table}[t!]
	\rowcolors{1}{white}{Gray}
	\centering
	\begin{threeparttable}
		\caption{A list of the tools used during the feature extraction process}
		\label{tbl_tools_feature_extraction}
		\begin{tabular}{p{5cm}p{1.5cm}L{1.5cm}p{6cm}}
			\toprule
			\multirow{2}{2cm}{Tool} & \multirow{2}{1.5cm}{Type} & \multirow{1}{1.5cm}{Extraction Technique} & \multirow{2}{4cm}{Attribution System} \\ 
			\addlinespace[1.5em]
			\midrule
			angr \cite{Shoshitaishvilietal2016} & \textit{ds} & \Circle & \cite{Alrabaeeetal2018a} \\ 
			BinComp \cite{Rahimianetal2015} & \textit{cp} & \Circle  & \cite{Alrabaeeetal2019a} \\
			BinShape \cite{Shiranietal2017} & \textit{o} & \Circle & \cite{Alrabaeeetal2018a} \\
			bjoern \cite{bjoern} & \textit{ds} & \Circle & \cite{Caliskanetal2018}\\         	
			Cuckoo Sandbox \cite{cuckoo} & \textit{s} & \CIRCLE & \cite{Rosenblumetal2011}, \cite{Rosenbergetal2018}, \cite{Haddadpajouhetal2020} \\ 
			Custom Android App & \textit{u, p} & \Circle & \cite{Kalgutkaretal2018}, \cite{Gonzalezetal2018} \\
			DECAF \cite{Hendersonetal2017} & \textit{s} & \CIRCLE & \cite{Hendrikse2017} \\         	
			Dyninst \cite{Dyninst} & \textit{ds}  & \Circle & \cite{Rosenblumetal2011}\tnote{1}, \cite{Meng2016}, \cite{Mengetal2017}, \cite{MengMiller2018}, \cite{Alrabaeeetal2019a}\tnote{1}, \cite{Alrabaeeetal2019b}\tnote{1} \\ 
			FLOSS \cite{flarefloss} & \textit{se} & \Circle & \cite{Laurenzaetal2018} \\ 
			FOSSIL \cite{Alrabaeeetal2018b} & \textit{o} & \Circle & \cite{Alrabaeeetal2018a} \\ 
			IDA Pro/Hex-Rays \cite{IDAPro} & \textit{ds, d} & \Circle & \cite{Alrabaeeetal2014}, \cite{Hendrikse2017}, \cite{Caliskanetal2018}, \cite{Alrabaeeetal2019a}, \cite{Alrabaeeetal2019b} \\ 
			Jakstab \cite{jakstab} & \textit{ds} & \Circle & \cite{Alrabaeeetal2019a}, \cite{Alrabaeeetal2019b} \\         	
			Manually & \textit{o} & \RIGHTcircle & \cite{Marquis-Boire2015}\\ 
			Netwide Assembler \cite{nasm} & \textit{ds} & \Circle & \cite{Caliskanetal2018}\\ 
			Nucleus \cite{Andriesseetal2017} & \textit{ds} & \Circle & \cite{Alrabaeeetal2019b} \\
			pefile \cite{pefile} & \textit{o} & \Circle & \cite{Laurenzaetal2018}, \cite{Alrabaeeetal2019a}, \cite{Alrabaeeetal2019b} \\ 
			radare2 \cite{radare} & \textit{ds} & \Circle & \cite{Caliskanetal2018}\\ 
			Unknown tool used & \textit{o} &\RIGHTcircle & \cite{Hongetal2018} \\ 
			UPX \cite{upx} & \textit{u} &\RIGHTcircle & \cite{Alrabaeeetal2019a}, \cite{Alrabaeeetal2019b} \\ 
			\bottomrule
		\end{tabular}
		\begin{tablenotes}
			\footnotesize
			\item[1] \cite{Rosenblumetal2011}, \cite{Alrabaeeetal2019a} and \cite{Alrabaeeetal2019b} use \textit{ParseAPI} which is now included within Dyninst \cite{Dyninst}.
			\item \textbf{Key:-} \Circle\tab[0.2cm]- Static Analysis \tab[0.2cm] \RIGHTcircle\tab[0.2cm]- Static and Dynamic Analysis \tab[0.2cm] \CIRCLE\tab[0.2cm]- Dynamic Analysis \tab[0.2cm] 
			\item \tab[0.7cm ] \textit{ds} \tab[0.2cm]- disassembler \tab[0.2cm] \textit{o} \tab[0.2cm]- other \tab[0.2cm] \textit{s}\tab[0.2cm] - sandbox \tab[0.2cm] \textit{u}\tab[0.2cm] - unpacker\tab[0.2cm] \textit{p} \tab[0.2cm]- parser \tab[0.2cm] \textit{d} \tab[0.2cm]- decompiler
			\item \tab[0.7cm]  \textit{se} \tab[0.2cm]- string extractor \tab[0.2cm] 
			\textit{cp} \tab[0.2cm]- compiler provenance \tab[0.2cm] 
		\end{tablenotes}
	\end{threeparttable}
\end{table}

\subsubsection{Feature Extraction Tools} All these categories require tools able to extract features from varying aspects of the binary. 
We collated all the tools used in the eighteen systems in Table \ref{tbl_tools_feature_extraction}. This allows us to assess the popularity of each tool and understand why some tools are used more than others. We note most of the tools used are for static extraction. We observe the most popular tool as Dyninst \cite{Dyninst} closely followed by IDA Pro/Hex-Rays \cite{IDAPro}. The reason Dyninst most likely edges IDA Pro is due to Dyninst being open source.
Five of the systems use multiple tools to extract different features \cite{Alrabaeeetal2018a,Alrabaeeetal2019a,Alrabaeeetal2019b,Caliskanetal2018,Rosenblumetal2011}, and this appears to be the best approach for extracting features from the five macro-categories we recommend in Section \ref{sec:style}.
Two unpackers, UPX \cite{upx} and a custom Android app, were used by \citet{Alrabaeeetal2019a,Alrabaeeetal2019b} and \citet{Kalgutkaretal2018,Gonzalezetal2018} respectively. This shows the lack of interest in applying the current state-of-the-art methods to the malware domain. Only two dynamic analysis tools were used in total. \citet{Rosenbergetal2017,Rosenbergetal2018} and \citet{Haddadpajouhetal2020} both use Cuckoo Sandbox \cite{cuckoo} and \citet{Hendrikse2017} uses DECAF \cite{Hendersonetal2017}. Unfortunately, the tool used by \citet{Hongetal2018} is undisclosed.

Overall, a total of nineteen tools were used. This shows there exists limited knowledge on whether extracting the same features via different tools affects the ability to capture authorship style.

\subsubsection{Feature Comparison} \label{sec:macro_features} We collated all the features from the eighteen systems and organized them into the five feature macro categories related to malware author style. In total, we collated 72 features. We structured the features by cross-referencing them against the systematization of the data modeling techniques from Section \ref{sec:data_modeling} and present the results in \Cref{tbl:features_1} and \Cref{tbl:features_2}. Where possible, we condensed any papers into single columns which used exactly the same features. We also include the column ``extraction techniques'' to indicate the programming analysis techniques required to extract each feature. Due to all the ``assembly'' features requiring only static analysis extraction techniques, we present the categorization from the multiple author works \cite{Meng2016,Mengetal2017, MengMiller2018} alongside the single author works. We include the column ``Authorship Problem'' in \Cref{tbl:features_2} for easier comparison across both the single and multiple author problem.

From this results, we remark there exists no favorable feature set for which the research field currently agrees upon. In fact, we recorded 42 unique features. In terms of feature extraction, researchers show a clear preference towards static analysis (36 features) and in terms of favorable macro-category then there exists a clear preference towards ``assembly language''.  In the following, we provide additional insight into the five macro-categories in terms of the application of these features for MAA. We use the macro-categories due to the vast number of features.

\begin{table*}[ht!]
	\footnotesize
	\centering
	\begin{threeparttable}
		\caption{State-of-the-art strings, implementation, infrastructure and decompiler features used in binary and malware authorship attribution research.}
		\label{tbl:features_1}
		\addtolength{\tabcolsep}{-1pt}
		\begin{tabular}[H]{C{6.5cm}C{0.4cm}C{0.4cm}C{0.45cm}C{0.4cm}C{0.45cm}C{0.4cm}C{0.4cm}C{0.4cm}C{0.4cm}C{0.4cm}C{0.4cm}C{0.4cm}}
			
			\toprule
			\addlinespace[6.25em]
			
			\cellcolor{white} & \cellcolor{white} & \multicolumn{7}{m{2.8cm}}{\tab[0.5cm]\rothead{-6.25}{Classifying}} & \multicolumn{1}{p{0.4cm}}{\tab[0.1cm]\subrothead{-9}{Clustering}} & \multicolumn{1}{p{0.4cm}}{\tab[0.1cm]\subrothead{-9.25}{Anomaly Detection}} & \multicolumn{2}{m{0.8cm}}{\tab[0.1cm]\rothead{-6.25}{Non ML}} \\ 
			\cmidrule(){3-13}
			
			\rowcol
			\multicolumn{1}{p{4cm}}{\multirow{-2}{*}{\textbf{String Features}}} & \rothead{-6}{Extraction Technique} & \cite{Caliskanetal2018} & \cite{Rosenbergetal2017}  & \cite{Hongetal2018} & \cite{Rosenbergetal2018} & \cite{Kalgutkaretal2018} & \cite{Alrabaeeetal2019a} &   \cite{Alrabaeeetal2019b} & \cite{Haddadpajouhetal2020} &\cite{Laurenzaetal2018}&\cite{Marquis-Boire2015}&  \cite{Alrabaeeetal2018a} \\ \midrule
			Artifact naming schemes/Algorithms &\RIGHTcircle&&& &&&&&&&\cmark& \\
			\rowcol
			C\&C Commands &\RIGHTcircle&&&&&&& & & & \cmark& \\ 
			Cuckoo Sandbox Report (Treated as Words) & \RIGHTcircle  & & \cmark& & \cmark & & & & &&&  \\
			\rowcol
			Encryption Keys&\RIGHTcircle&&&&&&& &&&\cmark& \\
			Errors&\Circle&&&&&&& &&&\cmark&\cmark \\
			\rowcol
			File Header & \RIGHTcircle & &  & & && & & \cmark & \cmark& &\cmark \\
			Function Names & \Circle & &  & & & &&&&& \cmark&\cmark \\
			\rowcol
			Grammar Mistakes&\Circle&&&&&&&&& &\cmark& \\
			MS-DOS Header & \Circle & & & & && &&&\cmark&&  \\
			\rowcol
			N-Grams (Words) & \Circle & \cmark & & & &  \cmark & \cmark && &&\cmark&  \\
			Optional Header & \Circle & & && & & && &\cmark&&  \\
			\rowcol
			Operating System& \RIGHTcircle &&&&& &&&&&&\cmark \\
			Programming Language Keywords & \Circle & \cmark& & & &&& && &&\cmark \\
			\rowcol
			Timestamp Formatting&\Circle&&&&& &&&&&\cmark& \\ \bottomrule
			
			\rowcol
			& & & & &  && & && & & \\
			
			\rowcol
			
			\multicolumn{1}{p{4cm}}{\multirow{-2}{*}{\textbf{Implementation Features}}} &   &   &   &  &  & & & & && & \\ \midrule
			
			Binary Data Directories &  \Circle &   & & & & &  &   & & \cmark&  &    \\ 
			\rowcol
			C\&C Parsing Implementation &  \RIGHTcircle&  &  & & & & && & &  \cmark&   \\
			Code Re-use &  \Circle&  &  &  & &  & & && &  \cmark&   \\
			\rowcol
			Compiler &  \Circle &  &  &  &  & & & &  & &  \cmark&  \cmark \\
			Configuration Techniques &  \RIGHTcircle&  &  &  & & & & &    &  &  \cmark&   \\
			\rowcol
			Constructor Design &  \RIGHTcircle&  &  &   &  & & & & &  &  \cmark&   \\ 
			Cyclometric Complexity &   \Circle &   &   &  & & & &    \cmark & &   &   &   \\
			\rowcol			
			Execution Traces &   \CIRCLE &   &   &  & & & &    \cmark &  & &   &   \\ 
			File Interactions Traits (Locations,  Modified,  \textit{etc}) &   \RIGHTcircle &    &    & \cmark & & & &   &   & & \cmark&  \cmark \\
			\rowcol
			Function Lengths &   \Circle &   &  & & & &    &   & & \cmark&  &  \cmark \\ 
			Multithreading Model (Use of Mutexes) &   \Circle &    &   & \cmark & & & & &  &   &  \cmark&    \\
			\rowcol
			Obfuscated String Statistics &   \Circle &   &  & & & &   &   & & \cmark&  \cmark&    \\
			Obfuscation Functions &   \RIGHTcircle &  &  & & & &   &  &  & & \cmark&   \\
			\rowcol
			Propagation Mechanisms &  \RIGHTcircle&  &  &  & & & &    &  & & \cmark&   \\ 
			Registry Keys &   \CIRCLE &    &   & \cmark & & & &   &   &  &  &  \\
			\rowcol
			System API Calls &   \RIGHTcircle &  &  &   \cmark &  & & & \cmark &   \cmark &   &  \cmark&  \cmark \\
			System/OS Version Determination technique &  \Circle&   && & & &  &  &  & & \cmark&   \\
			\rowcol
			Software Architecture \& Design &  \RIGHTcircle&  &  && & & &  &  & & \cmark&   \\
			Stealth and Evasion Techniques &  \RIGHTcircle&  &  && & & &   &  & & \cmark&   \\
			\rowcol
			Use of Global Variables&  \Circle &  &  && & & &  &  & & \cmark& \\ \bottomrule
			\rowcol
			& & & & & & && & & & & \\
			\rowcol
			\multicolumn{1}{p{4cm}}{\multirow{-2}{*}{\textbf{Infrastructure Features}}} &  & & & & & \\ \midrule
			
			DNS URLs & \RIGHTcircle & & & \cmark & & & & & & & \cmark&  \\
			\rowcol
			IP addresses (C\&C Servers)& \RIGHTcircle &  &&  \cmark & & && & & & \cmark&  \\
			Network Communication& \RIGHTcircle && & & &  & & & & & \cmark & \cmark \\
			\rowcol
			User Agent/Beaconing Style& \RIGHTcircle  && & & & & & & & & \cmark & \\ \bottomrule
			
			\rowcol
			& & & & & & && & & & & \\
			\rowcol
			\multicolumn{1}{p{4cm}}{\multirow{-2}{*}{\textbf{Decompiler Features}}} & & & & & & & & & & & & \\ \midrule
			Abstract Syntax Tree & \Circle & \cmark & & & & & & & & & & \\ \bottomrule
			
		\end{tabular}
		
		\begin{tablenotes}
			\item \textbf{Key:-} \Circle\tab[0.1cm]- Static Analysis \tab[0.1cm] \RIGHTcircle\tab[0.1cm]- Static and Dynamic Analysis \tab[0.1cm] \CIRCLE\tab[0.1cm]- Dynamic Analysis
		\end{tablenotes}
		
	\end{threeparttable}
\end{table*}

\begin{landscape}
	\begin{table*}[ht!]
		\centering
		\footnotesize
		\begin{threeparttable}
			\caption{State-of-the-art assembly features used in binary and malware authorship attribution research. All assembly features are extracted using static analysis.}
			\label{tbl:features_2}
			\addtolength{\tabcolsep}{-1pt}
			\begin{tabular}[H]{C{8cm}C{0.4cm}C{0.45cm}C{0.4cm}C{1.5cm}C{0.4cm}C{0.4cm}C{0.4cm}C{0.4cm}C{0.4cm}C{0.45cm}C{0.4cm}C{0.4cm}C{1cm}C{0.4cm}C{0.4cm}C{0.4cm}}
				
				\toprule
				\addlinespace[5.5em]
				
				\cellcolor{white} & \cellcolor{white} & \multicolumn{8}{m{3.2cm}}{\rothead{-6.75}{Classifying}} & \multicolumn{2}{p{0.8cm}}{\tab[0.5cm]\subrothead{-9.5}{Clustering}} & \multicolumn{1}{p{0.4cm}}{\tab[-0.1cm]\subrothead{-10.5}{\begin{minipage}{1.25cm}Anomaly Detection\end{minipage}}} & \multicolumn{1}{m{1cm}}{\tab[0.1cm]\subrothead{-10.5}{\begin{minipage}{1.5cm}Structured Prediction\end{minipage}}} & \multicolumn{3}{m{1.2cm}}{\tab[0.1cm]\rothead{-6.75}{Non ML}} \\ 
				\cmidrule{3-17}
				
				\rowcol
				
				\multicolumn{1}{p{4cm}}{\multirow{-2}{*}{\textbf{Assembly Features}}} & \rothead{-6}{Authorship Problem} & \cite{Rosenblumetal2011}  & \cite{Caliskanetal2018} &\cite{Meng2016,Mengetal2017,MengMiller2018}& \cite{Hendrikse2017}& \cite{Hongetal2018} & \cite{Gonzalezetal2018} & \cite{Alrabaeeetal2019a} & \cite{Alrabaeeetal2019b} & \cite{Rosenblumetal2011} & \cite{Haddadpajouhetal2020} &\cite{Laurenzaetal2018}&\cite{Mengetal2017,MengMiller2018}& \cite{Alrabaeeetal2014} &\cite{Marquis-Boire2015}&  \cite{Alrabaeeetal2018a} \\ \midrule
				
				Annotated Control Flow Graph & \Circle &  & & & & & & \cmark  & & & & & &  && \\
				\rowcol
				Backward Slices of Variables  & \CIRCLE & & & \cmark  &  & & & & && & & \cmark & &&  \\
				Block Catches Exceptions  & \CIRCLE & & & \cmark  & & & & & & && & \cmark  & &\cmark&  \\
				\rowcol				
				Block Position Within a Function CFG  & \CIRCLE & & & \cmark  & & & & & & & && \cmark & &&  \\
				Block Throws Exceptions  & \CIRCLE & & & \cmark  & & & & & & && & \cmark  & &\cmark&  \\
				\rowcol
				Byte Codes  & \Circle & & &   & & & & & & & \cmark &   & & &&  \\
				Call Graphlets & \Circle & \cmark & & & & & &  & & \cmark & &  & & \cmark &&\cmark \\
				\rowcol
				CFG Edge Types  & \CIRCLE & & & \cmark & & & & & &&  && \cmark  & &&  \\
				Constant Values  & \CIRCLE & & & \cmark  & & & & & & && &\cmark & &&\cmark \\
				\rowcol
				Control Flow Graph Edges  \& Node Unigrams & \Circle &  & \cmark &  & & &&  & & & & &  &&  &  \\
				Control Flow Graph Hashes & \Circle &  &  &  & \cmark & & & & && & & &  &&  \\
				\rowcol
				Data Flow Graph & \Circle &  & & & & & & \cmark  & & && & &  && \\
				Exact Syntax Template Library & \Circle & & & & & & & && & & & & \cmark &&  \\
				\rowcol
				Function (Opcode Chunks) & \Circle &  & & & & \cmark & & && & & & &  &&\cmark \\
				Function CFG Width  \& Depth & \CIRCLE && & \cmark  &  && & && & && \cmark  & &&  \\
				\rowcol
				Graphlets & \Circle & \cmark & & & &  & &  & & \cmark & &  & & \cmark &&  \\
				Idioms (Instructions) & \Circle & \cmark & & \cmark& & & & & & \cmark & && \cmark & \cmark &&  \\
				\rowcol
				Imports  \& Exports (Shared Libraries, Method Names)& \Circle & & & & & \cmark & & & & & &\cmark& & &&  \\
				Inexact Syntax Template Library & \Circle & &  & & & & & & & && & & \cmark &&  \\
				\rowcol
				Instruction Operand Sizes  \& Prefixes  & \CIRCLE & & & \cmark  & & & & & & & && \cmark & &&  \\ 
				Library Calls & \Circle & \cmark & & \cmark & & & & & & \cmark && & \cmark & \cmark &&\cmark \\
				\rowcol
				Loop Nesting Level  & \CIRCLE & & & \cmark &  & & & & & &&& \cmark & &&  \\
				Loop Size  & \CIRCLE & & & \cmark &  & & & & & & && \cmark & &&\cmark \\ 
				\rowcol
				N-Grams (Opcodes) & \Circle & \cmark & & \cmark &  & & \cmark & & \cmark & \cmark &&& \cmark & \cmark &&\cmark \\
				Number of Basic Blocks&\Circle&&&&&& & &&&&&&&&\cmark \\
				\rowcol
				Number of Input/Output/internal registers of a block  & \CIRCLE & & & \cmark & & & & && & && \cmark & &&  \\
				Number of Live Registers at Block Entry \& Exit  & \CIRCLE & & & \cmark & & & & & & &&& \cmark & &&  \\
				\rowcol
				Number of Used \& Defined Registers   & \CIRCLE & & & \cmark & & & & & && && \cmark & &&  \\
				Opcodes  & \Circle & & &  & & & & & & & \cmark & &  & &&  \\
				\rowcol
				Register Flow Graph & \Circle & & & & \cmark & & & & & &&& & \cmark &&  \\
				Stack Height Delta of the Block   & \CIRCLE & & & \cmark & & & & & & &&& \cmark & &&\cmark \\
				\rowcol
				Stack Memory Accesses  & \CIRCLE & & & \cmark &  & & && & &&& \cmark & &\cmark&\cmark \\
				Super Graphlets & \Circle & \cmark & & & & & &  & & \cmark & &  & & \cmark &&  \\ \bottomrule
				
			\end{tabular}
			
			\begin{tablenotes}
				\footnotesize
				\item \textbf{Key:-} \Circle \tab[0.1cm] Single Authorship Problem \tab[0.2cm] \CIRCLE \tab[0.1cm] Multiple Authorship Problem
			\end{tablenotes}
			
		\end{threeparttable}
	\end{table*}
\end{landscape}

\paragraph*{Strings.} These features capture any strings, artifacts and values within the malicious binary. The author influences any embedded \emph{strings} and so there exists a wealth of knowledge on the author to gain from extracting strings. For example, strings may infer the native language of the author and therefore their potential location. Any naming conventions for both functions and artifacts infer any author personality and choices. Other significant choices for the author which strings help infer include programming language, encryption techniques and error handling messages.

Malware authors understand how much information can be leaked from strings and therefore continue to research methods for either removing author style or changing them to imitate another author. Freely available tools such as packers, obfuscators and strippers allow any author to remove their author style from strings/constants. The simple task of adding false artifacts or function names shall change author style within strings. 
We note the special case of the works by \citet{Rosenbergetal2017,Rosenbergetal2018}, who convert the MAA problem into a Natural Language Processing (NLP) problem through the use of analyzing Cuckoo Sandbox reports \cite{cuckoo}.

\paragraph*{Implementation.} Features in this category describe author choice involving both malware design and execution. Predominately, researchers extract these features during dynamic analysis which makes them much harder to obfuscate and mask. For example, the approach the author takes to interact with the victim (\textit{e.g., propagation method}). Some of these features can be mimicked (\textit{e.g., toolchain process}) and this potentially allows authors to imitate other authors. If dynamic analysis fails, then analysts must rely on much harder and more manual techniques to identify implementation features. These features also change depending on both the malware authors' development environment and victim's system making it harder to automate across varying types of malware.

\paragraph*{Infrastructure.} We use this category to describe any feature which relates to specific infrastructure choices made by the author, \textit{e.g., choice of IP for command and control server}. If a threat actor reuses the same infrastructure, then it may offer an easy attribution decision. However, it might not be straightforward: for example, authors might attack other authors to use their infrastructure to imitate them or the author may loan out their infrastructure. We also expect sophisticated authors to change their infrastructure for each attack, or at the very least, mask identifiers such as IP addresses through methods such as IP spoofing or proxies.

\paragraph*{Assembly Language.} We collate any feature extracted from the assembly language representation of the binary. Researchers mainly extract them using static analysis which make them amenable to automation. These features focus on capturing instructions, control flow, data flow, external interactions and register flow. This can be either at the function level of the binary or much more fine-grained through the basic blocks of the program. Capturing both the program flow and more fine-grained features makes it harder for the author to modify them for adversarial purposes. The assembly language also presents an opportunity to feed it directly as a raw input into a Deep Neural Network (DNN) \cite{Mengetal2017}. 

A malware analyst relies on the state-of-the-art disassembler to maintain author style. Otherwise, the authorship attribution problem morphs into the binary similarity problem. Furthermore, there exist multiple methods to build \emph{basic blocks} and then \emph{control flow graphs} (CFG) to understand the flow of the program.  CFG include important program aspects such as \emph{error handling}, \emph{functions} and \emph{library and system interactions}. Even when built, graphs provide further complications as the problem of subgraph isomorphism remains an NP complete problem \cite{Cook1971}. Therefore, alternative representations must be sought. However, these alternatives then become approximations through statistical representations which increases the likelihood of losing authorship style. Finally, choices in the toolchain process (\textit{e.g.,} CFG flattening) present even more difficulties to overcome when building flow graphs.

\paragraph*{Decompiler.} This process attempts to recover the source code from the binary and remains an unsolved problem \cite{Robbins2017}. State-of-the-art decompilers such as IDA \cite{IDAPro} recover code which closely represents the original source code, especially when no optimization or other code modifications occurred during the toolchain process. Source code recovery allows researchers to extract author style features determined from \emph{source code authorship attribution}, \textit{e.g., abstract syntax trees} \cite{Caliskanetal2018}. However, these features rely heavily on the state-of-the-art decompilers and any binary modifications tend to highly impact the ability to recover them \cite{EmmerikWaddington2004}.

\section{Real-world Application of State-of-the-art Systems}\label{sec:discussion}

In this section, we provide an evaluation of the eighteen BAA systems identifying key findings and open research challenges (Section \ref{sec:keyfindings}) and research recommendations (Section \ref{sec:recommendations}). We present our results in \Cref{tbl:eval} and group the systems into the five data modeling techniques from Section \ref{sec:data_modeling}. Our systems evaluation consists of reviewing the efficacy and functionalities of the eighteen systems. This comparison considers the applicability of the current state-of-the-art techniques to malware binaries and enables us to set out the future research directions.
Here, we define \textit{Efficacy} as the accuracy of a system achieving its desired goal. We compare the efficacy of three experiments: (i) compiled source code, (ii) obfuscation, and (iii) malware on the largest datasets systematized in \Cref{tbl:datasets}.
For operational capability, we must consider the overall implementation of the system. Thus, we devise the following five categories to compare system \textit{functionality} (based on the availability and reproducibility of a system): 

\begin{itemize}
	\item[$\ominus$] \textbf{System currently not available.} We received no reply from our correspondence or the authors were unable to share the system.
	\item[$\oslash$] \textbf{System partially available.} We were able to locate part of the system online but core components were missing.
	\item[$\odot$] \textbf{System does not compile.} We attempted to modify the source code and install previous dependencies. However, this ultimately was not possible.
	\item[$\otimes$] \textbf{System contains errors at runtime.} If we managed to construct the system, we found errors occurred whilst attempting to run evaluation experiments which we were unable to patch. 
	\item[$\oplus$] \textbf{System completes.} The system was able to run a malware evaluation test.
\end{itemize}

In addition to this, we devise various categories for further comparison of the systems.  We provide indication of the \emph{ground truth} used to perform the experiment, namely source code, binaries or Android applications. We compare the systems by the addressed \emph{authorship problem (single or multiple)}, and the \emph{feature extraction techniques} used (static, dynamic or a combination of both). We also compare whether the researchers implemented \emph{parallelization} or \emph{cross-validation} evaluations, and whether they took into account \emph{toolchains} or \emph{shared libraries}. From our author style feature systematization (Section \ref{sec:features}), we compare the systems over the five macro-categories (\textit{i.e.}, \emph{strings}, \emph{implementation}, \emph{infrastructure}, \emph{assembly} and \emph{decompiler}). Finally, using the discussion in Section \ref{sec:adversary}, we explore whether any researchers consider \emph{adversarial} challenges and \emph{privacy} implications to their authorship attribution systems using the following categories we devised:

\noindent\textit{Adversarial:}
\begin{itemize}
\item[$\square$] Researchers do not consider any attacks
\item[$\boxdot$] Researchers consider unsophisticated attacks
\item[$\blacksquare$] Researchers consider sophisticated attacks
\end{itemize}
\textit{Privacy:}
\begin{itemize}
\item[$\square$] Researchers do not consider privacy implications  
\item[$\boxdot$] Researchers mention privacy implications 
\item[$\blacksquare$] Researchers discuss the privacy implications
\end{itemize}

\subsection{Key Findings and Open Research Challenges}\label{sec:keyfindings}
From the criteria above, and from the results shown in \Cref{tbl:eval}, we categorize our key findings as follows. 
First, we explore \emph{System Goal and Datasets} focusing on ground truth and authorship problem solved. Then, we examine the effect of \emph{Languages, Code Re-use and Toolchains} and \emph{Attribution Features and Extraction Methods} on BAA systems. Next, we examine the \emph{System Functionality} and in particular consider the importance of training, reproducibility of results and availability for the state-of-the art systems. Finally, we explore the open challenges with regards to \emph{System Efficacy} and \emph{Adversarial Considerations} as well contextualizing the implications of these systems on \emph{Privacy and Ethics}. 
 
\paragraph*{System Goal and Datasets.} We note the majority of the systems focus on the single authorship problem. All the systems use \emph{closed-world assumption} and in the majority of cases use classification systems. This makes them impractical for use in the ``wild''. \citet{Rosenblumetal2011} remains the only paper to partially consider the open world problem by training a model based on the closed world model. However, the datasets used to train, test and evaluate the systems contain minimal consistency especially for the datasets containing malware. Although the use of compiled source code repositories provides a ground truth, it brings extra complexities with the necessity to consider all toolchain possibilities. There exists no verifiable, publicly available and sufficiently large APT dataset which researchers can use to build MAA systems. All of the current attribution systems use datasets which are static, leading to a discontinuous learning model which is likely to experience concept drift. This is a wider problem within machine learning, deep learning and artificial intelligence models. \citet{Kolosnjajietal2018} show concept drift occurs within malware detection models based on unevolving datasets.

\begin{landscape}
	\begin{table*}
		\centering
		\vspace{-3em}
		\begin{threeparttable}
			\caption{Comparison of the Analyzed Systems between 2011 and 2019. Organized by data modeling technique and cross-referenced against ground truth, authorship problem, features, system efficacy, system functionality and adversarial considerations.}
			\label{tbl:eval}
			\begin{tabular}{p{3cm}C{0.75cm}C{0.75cm}L{0.5cm}L{0.2cm}L{0.2cm}L{0.2cm}L{0.2cm}L{0.2cm}L{0.2cm}L{0.2cm}L{0.2cm}L{0.2cm}L{0.2cm}L{0.2cm}C{2cm}C{1cm}C{2cm}C{0.4cm}C{0.3cm}C{0.3cm}}
				\toprule
				\addlinespace[0.3em]
				\rowcol
				& & &  & &   &  & &  &  & \multicolumn{5}{m{1cm}}{Features} & \multicolumn{3}{c}{Efficacy\tnote{d e}} & & &\\
				& \multirow{10.5}{*}{Paper} \cellcolor{white} & \multirow{10.5}{*}{Year} \cellcolor{white} & \subhead{0.3}{Original Groundtruth\tnote{a}} & \subhead{0.2}{Authorship Problem\tnote{b}} & \subhead{0.2}{Analysis\tnote{c}}  & \subhead{0.2}{Parallelization} & \subhead{0.2}{Cross-Validation} & \subhead{0.2}{Toolchains} & \subhead{0.2}{Shared Libraries} & \subhead{0.2}{Strings} & \subhead{0.2}{Implementation} & \subhead{0.2}{Infrastructure} & \subhead{0.2}{Assembly} & \subhead{0.2}{Decompiler} & \subhead{1}{Source Code (\%)} & \subhead{0.5}{Obfuscation (\%)} & \subhead{1.3}{Malware (\%)} & \subhead{0.4}{Functionality\tnote{f}} & \subhead{0.3}{Adversarial\tnote{g}} & \subhead{0.23}{Privacy\tnote{h}} \\ 
				\addlinespace[6em]
				\midrule
				& \cite{Rosenblumetal2011} & \citeyear{Rosenblumetal2011} & S & \Circle & \Circle &  & \cmark & &  &  &  &  & \cmark &  & 51 & F 58 & ACC 34 &  $\oslash$ & $\square$ & $\square$ \\
				\rowcol
				
				& \cite{Meng2016} & \citeyear{Meng2016} & S & \CIRCLE & \Circle &  & \cmark & & \cmark  &  &  &  & \cmark&  & 52 & $-$ & $-$ & $\ominus$ & $\square$ & $\square$ \\
				
				& \cite{Mengetal2017} & \citeyear{Mengetal2017}  & S & \CIRCLE & \Circle & \cmark & \cmark & & \cmark & \cmark &  &  & \cmark &  & 58 & $-$ & $-$ & $\ominus$ & $\square$ & $\square$ \\
				\rowcol
				& \cite{Rosenbergetal2017} & \citeyear{Rosenbergetal2017}& M &  \Circle & \CIRCLE &  & \cmark & &  & \cmark &  &  &  & & $-$ & $-$ & 94.6  & $\ominus$ & $\square$ & $\square$ \\
				
				& \cite{Hendrikse2017} & \citeyear{Hendrikse2017} & S & \Circle & \RIGHTcircle &  & \cmark &  \cmark & \cmark  &  &  & & \cmark &  & 95.3 & 94.1  & $-$ & $\ominus$ & $\boxdot$ & $\boxdot$\\
				\rowcol
				& \cite{Caliskanetal2018} & \citeyear{Caliskanetal2018} & S & \Circle & \Circle &  & \cmark & & & \cmark &  &  & \cmark & \cmark & 83 & 88 & ACC 70  & $\odot$ & $\boxdot$ & $\blacksquare$\\
				
				& \cite{MengMiller2018} & \citeyear{MengMiller2018} & S & \CIRCLE & \Circle & \cmark &  \cmark  & \cmark & \cmark & & &  & \cmark  &  & 71 & $-$ & $-$ & $\ominus$ & $\square$ & $\square$\\
				\rowcol
				& \cite{Hongetal2018} & \citeyear{Hongetal2018} & M & \Circle & \RIGHTcircle &  & \cmark & &  &  & \cmark & \cmark  & \cmark &  & $-$ & $-$ & AF 88.2 & $\ominus$ & $\square$ & $\square$ \\
				
				& \cite{Rosenbergetal2018} & \citeyear{Rosenbergetal2018} & M & \Circle & \CIRCLE &  & \cmark & &  & \cmark  &  &  &  &  & $-$ & $-$ & 99.75 & $\ominus$ & $\square$ & $\square$\\
				\rowcol
				& \cite{Kalgutkaretal2018} & \citeyear{Kalgutkaretal2018} & A & \Circle & \Circle &  & \cmark  & &  & \cmark &  &  &  &  & 98 & 77 & 96 & $\ominus$ & $\boxdot$ & $\square$\\
				
				& \cite{Gonzalezetal2018} & \citeyear{Gonzalezetal2018} &  A & \Circle & \Circle &  &  \cmark & &  &  & &  & \cmark &  & 86.74 & $-$ & 66.92 & $\ominus$ & $\square$ & $\square$\\
				\rowcol
				& \cite{Alrabaeeetal2019a} & \citeyear{Alrabaeeetal2019a} &  S & \Circle & \RIGHTcircle &  &  \cmark & \cmark & \cmark & \cmark & &  & \cmark &  & F 94 & $-$ & CC 96.9 & $\ominus$ & $\square$ & $\square$\\
				
				& \cite{Alrabaeeetal2019b} & \citeyear{Alrabaeeetal2019b} &  S & \Circle & \Circle &  &  \cmark & \cmark & \cmark &  & \cmark &  & \cmark  &  & P 84 & $-$ & P 45 & $\ominus$ & $\square$ & $\boxdot$\\
				\rowcol
				\multirow{-15}{*}{Classifying} & \cite{Alrabaeeetal2019b} & \citeyear{Alrabaeeetal2019b} &  S & \CIRCLE & \Circle &  &  \cmark & \cmark & \cmark &  & \cmark &  & \cmark &  & P 89 & $-$ & $-$ & $\ominus$ & $\square$ & $\boxdot$\\ \midrule

				& \cite{Rosenblumetal2011}  & \citeyear{Rosenblumetal2011} & S & \Circle & \Circle &  & \cmark & &  &  &  &  & \cmark &  & AMI 45.6 & $-$ & $-$ &  $\oslash$ & $\square$ & $\square$\\
				\multirow{-2}{*}{Clustering} & \cite{Haddadpajouhetal2020}  & \citeyear{Haddadpajouhetal2020} & M & \Circle & \CIRCLE &  &  & &  & \cmark  & \cmark &  & \cmark &  & $-$ & $-$ & $95$ &  $\ominus$ & $\square$ & $\square$\\ \midrule			
				
				\rowcol
				Anomaly Detection & \multirow{1}{*}{\cite{Laurenzaetal2018}} & \multirow{1}{*}{\citeyear{Laurenzaetal2018}} & \multirow{1}{*}{M}  & \multirow{1}{*}{\tab[-0.1cm]\Circle} & \multirow{1}{*}{\tab[-0.1cm]\Circle} &  &  \multirow{1}{*}{\cmark} & &  & \multirow{1}{*}{\cmark} & \multirow{1}{*}{\cmark} & & \multirow{1}{*}{\cmark} &  & \multirow{1}{*}{$-$} & \multirow{1}{*}{$-$} & \multirow{1}{*}{98} & \multirow{1}{*}{$\otimes$} & \multirow{1}{*}{$\square$} & \multirow{1}{*}{$\square$}\\ \midrule
				
				& \cite{Mengetal2017} & \citeyear{Mengetal2017} & S & \CIRCLE & \Circle  & \cmark & \cmark & & \cmark & \cmark &  &  & \cmark &  & 65 & $-$ & $-$ & $\ominus$ & $\square$ & $\square$\\
				\rowcol
				\multirow{-2}{1.5cm}{Structured Prediction} & \cite{MengMiller2018} & \citeyear{MengMiller2018} & S & \CIRCLE & \Circle & \cmark &  \cmark  & \cmark & \cmark & &  &  & \cmark  &  & $\times$ & $-$ & $-$ & $\ominus$ & $\square$ & $\square$\\ \midrule
				
				& \cite{Alrabaeeetal2014} & \citeyear{Alrabaeeetal2014} & S & \Circle & \Circle &  &  & & \cmark &  &  &  &   \cmark &  & 84 & F 25  & ACC 69.75 & $\ominus$ & $\square$ & $\square$ \\
				\rowcol
				& \cite{Marquis-Boire2015} & \citeyear{Marquis-Boire2015} & M & \Circle & \RIGHTcircle & &  & & & \cmark  & \cmark & \cmark  &  &  & \emph{n/a} & \emph{n/a} & \emph{n/a} & \emph{n/a} & $\square$ & $\square$\\
				
				\multirow{-3}{*}{Non-ML} & \cite{Alrabaeeetal2018a} & \citeyear{Alrabaeeetal2018a} & S & \Circle & \Circle  &  &  & \cmark & \cmark  & \cmark  & \cmark  & \cmark   & \cmark  &  & P 49 & P 95 & Re 68 &  $\otimes$ & $\boxdot$ & $\square$\\
				
				\bottomrule
			\end{tabular}
			\vspace{-1.35em}
			\begin{tablenotes}
				\footnotesize
				\begin{multicols}{2}
					\item[a] S - Source Code \tab[0.2cm] M - Malware \tab[0.2cm] A - Android Applications
					\item[b] \Circle \tab[0.1cm] Single Authorship Problem \tab[0.2cm] \CIRCLE \tab[0.1cm] Multiple Authorship Problem
					\item[c] \Circle \tab[0.1cm] Static Analysis \tab[0.2cm] \RIGHTcircle \tab[0.1cm] Static and Dynamic Analysis \tab[0.2cm] \CIRCLE \tab[0.1cm] Dynamic Analysis
					\item[d]  $\times$ \tab[0.1cm] Experiment Incomplete \tab[0.2cm] $-$ \tab[0.1cm] No Experiment Considered \tab[0.2cm] \emph{na} \tab[0.1cm] Not Applicable
					\item[e] All accuracy unless precedes with:	F - $F_{1}$ measure [except for Alrabaee et al. [7] who define and use $F_{0.5}$] AF - Average $F_{1}$ score \tab[0.2cm] AMI - Adjusted Mutual Information \tab[0.2cm] ACC - Average Correctly Clustered \tab[0.2cm] P - Precision \tab[0.2cm] CC - Correctly Clustered 
					\item Re - The average accuracy in relation to a malware analysis report
					\item[f] $\ominus$\tab[0.2cm] System Not Available $\oslash$\tab[0.2cm] System Partially Available $\odot$\tab[0.2cm] System Does Not Compile \tab[0.1cm] $\otimes$\tab[0.2cm] System Contains Errors At Runtime \emph{n/a} \tab[0.1cm] Not Applicable
					\item[g] $\square$ \tab[0.1cm] Researchers do not consider any attacks  \tab[0.1cm] $\boxdot$ \tab[0.1cm] Researchers consider unsophisticated attacks \tab[0.1cm] $\blacksquare$ \tab[0.1cm] Researchers consider sophisticated attacks
					\item[h] $\square$ \tab[0.1cm] Researchers do not consider privacy implications  \tab[0.1cm] $\boxdot$ \tab[0.1cm] Researchers mention privacy implications \tab[0.1cm] $\blacksquare$ \tab[0.1cm] Researchers discuss the privacy implications
				\end{multicols}
			\end{tablenotes}
		\end{threeparttable}
	\end{table*}
\end{landscape}

Malware development follows a similar agile work flow process to benign software development where multiple authors collaborate \cite{Callejaetal2016}, as in the recent GandCrab ransomware campaign \cite{Herzog2018}. However, there exists limited research exploring multiple authorship within MAA. Even in BAA, only four out of the eighteen papers consider multiple authorship for a program. From the four multiple author focused papers, there still remain research gaps such as applying these techniques to malware. However, there exist many challenges with this approach. This includes overcoming both obfuscation and packing techniques. Additionally, it remains unclear whether the features they use help with clustering multiple malware authors.

\paragraph*{Languages, Code Re-use and Toolchains.} Code re-use from other software and libraries impact attribution systems and this can lead to the incorrect author attributed. We observe only eight systems attempt to account for the effect of shared libraries on authorship style. Previous works all attempt to remove standard libraries from the binaries before extracting author features \cite{Meng2016,Mengetal2017,MengMiller2018,Alrabaeeetal2014,Alrabaeeetal2018a, Alrabaeeetal2019a, Alrabaeeetal2019b, Hendrikse2017}. These works only focus on removing C/C++ libraries due to their datasets containing binaries compiled from C/C++ source code. In fact, none of the state-of-the-art systems consider any other programming languages. 

However, authors write malware in multiple languages \cite{Callejaetal2019} and thus a programming language gap exists when it comes to identifying the malware author. Therefore, we believe using systems trained only on compiled source code datasets to label unknown malware hinders a MAA system. Further issues exist if the programmer adheres to language standards where a strict format must be followed, \textit{e.g,} the style guide for Python (PEP 8 \cite{Vanrossumetal2001}). Standards are most likely to significantly reduce the amount of author style within a program as everyone will produce similar looking code. However, the speed of malware development must match the speed at which it requires deploying\footnote{The window of deployment depends on the availability of a vulnerability patch.} and this determines the likelihood of a malware author following standards. In any case, future research should consider features which are robust against any standards to prevent this becoming an attack method to the attribution systems themselves.

The re-use of code within benign programs is common practice and malware development is no different. Within malware, there exists a lot of code re-use from both open and closed sources due to the pressure of beating vulnerability patching or meeting the demands of cyber warfare to complete mission objectives. Code re-use can be both helpful and unhelpful. In fact, a lot of code reuse from other authors contaminates samples and leads to an even smaller dataset to learn author style. For example, if someone leaks the source code then this quickly leads to multiple copycat attackers. On the other hand, code reuse of the actual malware author helps identify malware written by the same authors \cite{RosenbergBeek2018}.
Only three papers within the current research consider the effect of toolchains on their attribution system \cite{MengMiller2018,Hendrikse2017,Alrabaeeetal2018a}, meaning there still exist questions regarding the impact of compilers on author style. However, if we consider a malware dataset then the choice of compiler is predefined by the author and so by default we automatically would train upon a dataset which potentially used various compilers. This may answer why using a model trained on compiled source code provides limited aid when applying it to malware.

\paragraph*{Features, Extraction and Style.} We observe \textit{strings} and \textit{assembly language} as the two most popular feature macro-categories for author style and this also correlates with static analysis as the most popular technique to extract features. These popular macro-categories omit key malware specific features and traits which experts tend to discover among APT author style.
The common extraction process used involves static analysis, likely due to the ``quickness'' it provides over dynamic analysis. Furthermore, there exist multiple tools for binary analysis which achieve similar tasks. This leads to further research questions surrounding the effect of extraction tools on author style.

There exists limited research around finding malware author style. The goals of benign software programmers clearly differ to malicious software programmers and yet most of the research focuses on only the benign stylometry approach of lexical, syntactic and semantic features on assembly language and these methods ignore malware specific features. In the case of multiple authors, the state of the art mainly identifies fine-grained features (\textit{e.g.,} basic block exception handling) \cite{Meng2016,Mengetal2017, MengMiller2018, Alrabaeeetal2019b} and this differs to the features identified by \citet{Marquis-Boire2015} for linking malware authors (\textit{e.g.,} languages used, command and control server setups and obfuscation techniques used).

\paragraph*{System Functionality.} The majority of systems we tested, retrained their systems to perform each of the evaluation tests\footnote{Unfortunately, continuously retraining your system to account for new discoveries in the ``wild'' remains a resource intensive task.}. The system which showed higher performance capability in some cases required a training time of a week \cite{Mengetal2017} with the most likely explanation of using no parallelization techniques. 
After spending a considerable amount of time and effort, none of the eighteen systems we tested fell into the ``System completes'' category and this provides the main reason for a shortage of further research within this field. Although the published results show promise, the lack of consistency with the evaluation metrics makes it hard to validate the results without further testing. 

\paragraph*{System Efficacy.} Not all the papers performed experiments using source code, obfuscation or malware experiments\footnote{We utilized the results of the survey by \cite{Alrabaeeetal2017} to incorporate the obfuscation and malware experiments using the systems from \cite{Rosenblumetal2011,Caliskanetal2018, Alrabaeeetal2014}. However, \cite{Alrabaeeetal2017} omit the accuracy results for these experiments and instead use $F_{0.5}$ for the F-measure as they claim the systems in \cite{Rosenblumetal2011,Caliskanetal2018, Alrabaeeetal2014} are  extra sensitive to false positives.} and in some unique cases the system takes too long to complete \cite{MengMiller2018} or the method used is not applicable \cite{Marquis-Boire2015}. In terms of presented results we gathered, only 19 out of 34 used the accuracy metric. Therefore, we considered an alternative metric for the remaining 15 results. This highlights the lack of consistency on evaluation metrics across the field. Let us examine the three types of experiments:
\begin{itemize}
	\item \textbf{Source Code.} The results vary considerably and this makes it difficult to compare the systems. The later systems seem to perform better. This appears to be down to the progress of extraction techniques which allow researchers to remove some external noise (\textit{e.g.} system libraries) from the binaries. %
	\item \textbf{Obfuscation.} From the very few experiments, it remains impossible to tell whether the problem is solved due to the inconsistency of results. The result by \citet{Hendrikse2017} appears the most promising. This is due to the thoroughness of obfuscation techniques considered, and even though they considered the fewest number of features, the prominent difference to the other systems is the inclusion of dynamic features.
	\item \textbf{Malware.} Comparing the malware experiments is much harder, as the goals of the systems differ slightly. The datasets used were also considerably smaller, and the researchers undertook considerable efforts to clean the datasets. It is these reasons which explain the considerably high accuracy attained. This is not necessarily bad as it could help malware analysts examine a small subset of malware which they believe originate from the same author. Even if we were to consider much larger and dirtier datasets, then the state-of-the-art systems remain unlikely to produces the same levels of accuracy.
\end{itemize}

We note the inconsistency of datasets encumbers the comparison of the systems.  A prime example of this inconsistency is \citet{Caliskanetal2018}, who report better efficacy on obfuscation than \citet{Alrabaeeetal2017} despite appearing to use the same method for obfuscation experiments. Even though both papers report different metrics, we state the reasons we think there exists a higher accuracy in the later results. Firstly, we believe \citet{Alrabaeeetal2017} used an older version of the system from \cite{Caliskanetal2018} as they published their paper first. Secondly and most importantly, they both use different datasets.

\paragraph*{Adversarial Considerations.} From the table, only four\footnote{The obfuscations experiments for systems \cite{Rosenblumetal2011} and \cite{Alrabaeeetal2014} were computed in the survey by \citet{Alrabaeeetal2017}.} of the eighteen systems \cite{Caliskanetal2018,Hendrikse2017, Kalgutkaretal2018, Alrabaeeetal2018a} considered basic attacks, \textit{e.g,} obfuscation. %
This highlights the lack of adversarial considerations towards any binary attribution system. Even those researchers who implemented unsophisticated attacks (\textit{e.g.,} obfuscation) on their systems, reported an increase in the amount of manual assistance needed to de-obfuscate the binaries. This meant the systems became more semi-automated. Out of all the single authorship methods, \citet{Hendrikse2017} provides the most comprehensive evaluations using readily available obfuscation tools which range from very easy to hard techniques. However, their attribution system uses the fewest amount of features which opens itself to targeted and untargeted attacks. This is because their system uses fewer features than \citet{Caliskanetal2018} and \citet{Mengetal2018} show the binary attribution system created by \citet{Caliskanetal2018} is open to both: targeted and untargeted attacks. 
\citet{Mengetal2018} extend the attacks by \citet{CarliniWagner2017} designed for DNNs trained for image labeling. \citet{Mengetal2018} generated a method to modify the feature vector and the binary. When they modify the binary they ensure the binary still executes which is a fundamental requirement for a successful binary modification attack. We predict this method of attack works for all the other single author systems too. %
Therefore, the majority of single author state-of-the-art systems remain open to both unsophisticated and sophisticated attacks.  

\paragraph*{Privacy and Ethics.} The majority of systems use author style features developed from benign source code author identification rather than focusing on malicious author styles. This means these systems and techniques can be used to identify benign software developers who might create programs to avoid detection in nations which prevent freedom of speech. Furthermore, these authors may have previously submitted software to the benign sources used by many of the systems. The authors may be unaware of researchers using their software. This not only violates their privacy rights but this raises ethical questions surrounding the further use of the benign datasets.

\subsection{Recommendations}\label{sec:recommendations}
\paragraph*{Real-world Application.} None of the MAA systems we reviewed appear immediately ready for implementation in the ``wild''. There exists a lack of sufficient details to replicate the systems. Anyone wishing to join this research field must start from scratch and redo the majority of previous work. Furthermore, limited results on malware exist meaning it is unknown whether the current techniques are effective for real-world use. Additionally, most systems require intense manual analysis and significant training times further showing these systems are unready for operational deployment.
\paragraph*{Privacy.} Although these systems are aimed at detecting malicious authors, they can be used to detect benign software users and this raises privacy concerns. This provides further evidence future research must focus purely on malicious author styles. Few of these works consider the privacy and anonymity implication of the developed tools. Therefore, we believe MAA systems should also be tested in other contexts than that of malware written by a threat actor to measure their efficacy and impact in benign scenarios.
\paragraph*{Adversarial Approach.} None of the analyzed papers consider sophisticated adversarial testing. We suggest any MAA system must undergo adversarial testing before deployment. In particular, it must show robustness to sophisticated attacks like the one described by \citet{Mengetal2018} which we predict works for all current single author binary attribution systems. There also exists no research into adversarial attacks on multiple authorship attribution systems. In the future, we predict all APT malware authors shall implement sophisticated attacks to remain anonymous and avoid law enforcement.
\paragraph*{Datasets.} In general, there lacks both a consistency of performance metrics and datasets used across the research field. Systems which trained upon source code and were then used to identify malware author performed worse than those systems which originally trained upon malware. The datasets used played a pivotal role in these systems and most of them lacked a variety of programming languages or ability to cope with the effect of shared libraries and compilers. We hope the creation of our APT malware dataset in Section \ref{sec:new_dataset} allows a fair comparison among future systems. 
\paragraph*{Multiple Authors.} The single author assumption fundamentally hinders the ability to determine the author of binaries developed by multiple authors (\emph{agile software development}), especially as the commercialized malware industry uses agile work flow methods to speed up the development process to both increase profits and beat vulnerability discovery time. Being able to see if authors are used across multiple malware development projects shall provide insight within the malware development industry and introduce a new method of tracking malicious threats, especially APT groups. To further aide this we suggest all future work should adopt our approach of considering features from the five feature macro-categories of: \emph{strings}, \emph{implementation}, \emph{infrastructure}, \emph{assembly language} and \emph{decompiler}. This allows for all aspects of malware author styles to be captured.
\section{APTClass: Creation of an APT Malware Dataset}\label{sec:new_dataset}

From our discussion in Section \ref{sec:datasets} on datasets, we deemed it a high priority to ensure there exists a sufficiently large and diverse dataset accessible to research for use in discovering malware authorship style and creating malware authorship identification systems.  In this section, we set out how we created \textbf{APTClass}, a meta-information dataset consisting of 15,660 labeled malware samples. Our overall approach follows a similar method to \citet{aptdataset}: we gather a large amount of open-source intelligence (OSINT) and then we perform preprocessing on the data before extracting information. In addition, we propose a novel method for label identification and extraction to solve the issues discovered in Section \ref{sec:malwaredatasets} and because of our focus on labeling we only extract malware hashes. This can be extended to include \textit{URLs, IP Addresses, or Tactics, Techniques and Procedures} as shown by previous works \cite{Liaoetal16, Zhaoetal20}. Our novel label extraction method uses a matching algorithm which combs the OSINT in a systematic process to match against a list of 1,532 APT group names.  We describe this process in detail in Section \ref{sec:aptdataset-method}.

\subsection{Method}\label{sec:aptdataset-method}
APTClass follows five steps: (i) create a list of APT groups and group them by alleged nation; (ii) gather OSINT, mainly PDF reports of attack campaigns; (iii) extract hashes and label from the gathered intelligence; (iv) clean the dataset by removing duplicate malware hashes and use VirusTotal \cite{virustotal} to verify the legitimacy of the samples gathered in step (iii); and finally (v) filtering for executable binaries. For the purpose of this dataset, APTClass considers executable files as ELF, Windows 32 EXE, Windows 16 EXE, Windows Installation file and Windows DLL.

\begin{table*}[h!]
	\rowcolors{1}{white}{Gray}
	\centering
	\caption{List of sources used for creating a consistent list of APT labels.}\label{tbl:aptclass-sources}
	\begin{tabular}{lL{8cm}}
		\toprule
		\rowcolor{white}
		Source Name &  Last Updated \\
		\midrule
		MISP \cite{misp} & October 2020 \\
		APT Operation Tracker \cite{aptspreadsheet} & October 2020 \\
		MITRE ATT\&CK \cite{mitre} & October 2020 \\
		sapphirex00 \cite{sapphirex00} & Nov 2018 \\
		Thailand CERT \cite{thaicert} & October 2020 \\
		Council on Foreign Relations \cite{cfr} & October 2020 \\
		\bottomrule
	\end{tabular}
\end{table*}

\subsubsection{Creating a consistent list of APT labels} \label{sec:aptclassi}
To overcome the issues of multiple aliases introduced by various analysts, APTClass treats each name as a unique group. Although this initially inflates the number of groups and introduces some duplication (\textit{e.g., group 123} and \textit{group123} are listed separately), we believe this to be the correct approach as often analysts cannot reach a consensus regarding groups and may use different names within OSINT when referring to the same group.
APTClass still captures any nation link for a group. From our experience, analysts tend to have a higher confidence on linking groups to nations. APTClass also records whether a group is linked to multiple nations to account for mis-attribution. APTClass extracts the nation and group names from six sources in \Cref{tbl:aptclass-sources} using the process set out in \Cref{alg:apt_list}. Essentially, APTClass extracts from the sources a \textit{dictionary} with \emph{nations} as \textit{keys} and \emph{a list of group names} as \textit{values}. APTClass then standardizes the names and removes duplicates over the six dictionaries. This approach identified 1,532 names. We are aware there exists duplication among sources, however, this helps further validate the list of names as well as increase the varying aliases for each group.

\begin{algorithm}[h!]
	\SetAlgoLined
	\SetKwInput{KwInput}{Input}                %
	\SetKwInput{KwOutput}{Output}              %
	\DontPrintSemicolon
	
	\KwInput{$sources\_list$}
	\KwOutput{$ final\_list $}
	
	\SetKwFunction{FMain}{Main}
	\SetKwFunction{FDuplicates}{remove\_duplicates}

	\SetKwProg{Fn}{Function}{:}{\KwRet}
	\Fn{\FMain}{
		\For{source in sources\_list}{
			dictionary($ nation $ : $ group\_name\_list $) = extract\_nation\_and\_names($ source $)\; \tcp*{returns a dictionary, with nations as keys and list of group names as values}
			\For{each nation}{
				$ group\_name\_list $ = standardize($ group\_name\_list $)\; \tcp*{removes punctuation and converts to lowercase}
				$ group\_name\_list $ = remove\_duplicates($ group\_name\_list $)\; \tcp*{removes any duplicates from the list of group names}
			}
			\For{name in group\_name\_list}{
				\If{(nation, name) not in final\_list}{
					$final\_list$.append($(nation, name)$)\;
				}
			}	
		}
		$final\_list$ = group\_nations($final\_list$)\;  \tcp*{joins together groups from the same nations}
		\KwRet $final\_list$\;
	}
	\caption{Creating list of APT names}\label{alg:apt_list}
\end{algorithm}

\subsubsection{Gathering open-source threat intelligence} \label{sec:aptclassii}
We performed an extensive search on GitHub for trustworthy repositories containing any OSINT information. In particular, we focused on repositories storing (i) reports (typically PDF files), (ii) indicator of compromises (IoC), and (iii) YARA rules. We chose GitHub as the majority of OSINT is shared on the platform from other researchers collecting their own repositories of intelligence. We wanted to collate as many files as possible to ensure we maximized the number of malware hashes.

\subsubsection{Extracting hashes and labels} 
We are aware of many OSINT parsers, however, these just extract the indicators of compromise \cite{iocparser, observable} or try to gather tactics and techniques for groups \cite{legoyetal2020} without extracting the most important piece of information for our own purpose: APT labels and hashes. Thus, we required a new approach to gather a likely label for the malware hash. APTClass provides a fine-grained approach to extracting the label. We set out this technique in \Cref{fig:extraction} and describe the process below:

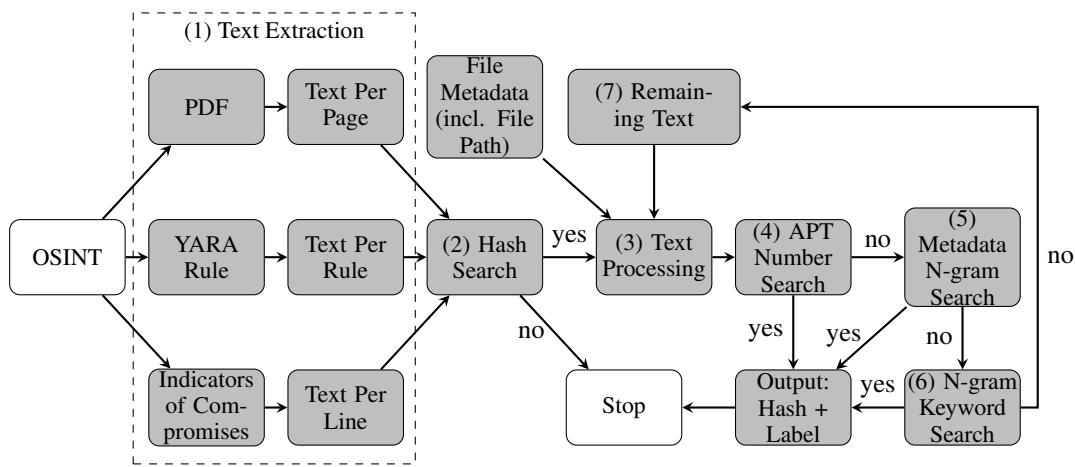
\begin{figure}[ht!]
	\begin{minipage}[t][6cm]{\columnwidth}
		\begin{center}
			\begin{tikzpicture}[node distance = 0mm]
				\tikzstyle{process} = [rectangle, rounded corners, minimum width=1cm, minimum height=1cm,text centered, draw=black, fill=gray!50, font=\footnotesize, inner sep=0.5pt, text width=1.5cm]
				\tikzstyle{sink} = [rectangle, rounded corners, minimum width=1cm, minimum height=1cm,text centered, draw=black, fill=white, font=\footnotesize, inner sep=0.5pt, text width=1.5cm]
				\tikzstyle{arrow} = [thick,->,>=stealth]
				\node (BG1) at (-16,5) [draw=black, dashed,minimum width=3.75cm,minimum height=6cm, xshift=2.75cm, yshift=0.5cm] {};
				\node at (-13.25,8.25)[font=\small]{(1) Text Extraction};
				\node (start) [sink] at (-16,5.25){OSINT};
				\node (extract1a) [process, right of=start, xshift=1.85cm, yshift=2cm] {PDF};
				\node (extract1b) [process, right of=extract1a, xshift=1.85cm] {Text Per Page};
				\node (extract2a) [process, right of=start, xshift=1.85cm] {YARA Rule};
				\node (extract2b) [process, right of=extract2a, xshift=1.85cm] {Text Per Rule};				
				\node (extract3a) [process, right of=start, xshift=1.85cm, yshift=-2cm] {Indicators of Compromises};
				\node (extract3b) [process, right of=extract3a, xshift=1.85cm] {Text Per Line};
				\node (hash) [process, right of=extract2b, xshift=1.85cm] {(2) Hash Search};
				\node (filepath) [process, right of=hash, yshift=2cm] {File Metadata (incl. File Path)};	
				\node (proc) [process, right of=hash, xshift=2.25cm] {(3) Text Processing};
				\node (label1) [process, right of=proc, xshift=1.85cm] {(4) APT Number Search};
				\node (end) [process, right of=label1, yshift=-2cm] {Output: Hash + Label};
				\node (label2) [process, right of=label1, xshift=2.25cm] {(5) Metadata N-gram Search};
				\node (label3) [process, right of=label2, yshift=-2cm] {(6) N-gram Keyword Search};		
				\node (label4) [process, right of=proc, yshift=2cm, minimum width=1.5cm, text width=2.25cm] {(7) Remaining Text};
				\node (stop) [sink, right of=hash, xshift=1.85cm, yshift=-2cm] {Stop};									
				\draw [arrow] (start) -- (extract1a);
				\draw [arrow] (start) -- (extract2a);
				\draw [arrow] (start) -- (extract3a);
				\draw [arrow] (extract1a) -- (extract1b);
				\draw [arrow] (extract2a) -- (extract2b);
				\draw [arrow] (extract3a) -- (extract3b);
				\draw [arrow] (extract1b) -- (hash);
				\draw [arrow] (extract2b) -- (hash);
				\draw [arrow] (extract3b) -- (hash);
				\draw [arrow] (hash) -- node[anchor=south] {yes} (proc);
				\draw [arrow] (filepath) -- (proc);
				\draw [arrow] (hash) -- node[anchor=east] {no} (stop);
				\draw [arrow] (proc) -- (label1);
				\draw [arrow] (label1) -- node[anchor=south] {no} (label2);
				\draw [arrow] (label1) -- node[anchor=east] {yes} (end);
				\draw [arrow] (label2) -- node[anchor=east] {no} (label3);
				\draw [arrow] (label2) -- node[anchor=east] {yes} (end);
				\draw [arrow] (label3) -- node[anchor=south] {yes} (end);
				\draw [arrow] (label3) |- + (1,0) |- node[anchor=west,yshift=-2cm] {no} (label4);
				\draw [arrow] (label4) -- (proc);
				\draw [arrow] (end) -- (stop);
			\end{tikzpicture}			
		\end{center}
	\end{minipage}
	\caption{A high-level view of the extraction process for APTClass.}
	\label{fig:extraction}
\end{figure}

\begin{enumerate}
	\item \textbf{Text Extraction:} APTClass extracts the text per page of PDF reports, text per YARA rule and text per line of IoC files. This allows APTClass to try and identify the best possible label closest to the hash.
	\item \textbf{Hash Search:} We perform a regular expression search on the extracted text for any MD5, SHA1, SHA256 or SHA512 hashes.
	\item \textbf{Text Processing:} APTClass removes punctuation, stop words and hashes from the text. The stop words consist of stop words from NLTK \cite{nltk}, spaCy \cite{spacy} and gensim \cite{gensim} as well as any cyber words in the dictionary created by Bishop Fox \cite{bishopfox} and words previously determined ``noise'' from running APTClass multiple times.
	\item \textbf{APT Number Search:} APTClass performs an extensive search against the APT label list looking for a match with either:
	\begin{center}
		\begin{minipage}{0.4\textwidth}
			\begin{itemize}
				\item APT$<$number$>$,
				\item APT-C-$<$number$>$,
				\item ATK$<$number$>$,
				\item SIG$<$number$>$ or 
				\item FIN$<$number$>$
			\end{itemize}
		\end{minipage}
	\end{center}
	We do this as these labels tend to be extremely popular labels among analysts. APTClass only uses this as a label if there is a clear majority within the matches. APTClass also designates this match as the label when there is no further match against the APT label list created in Section \ref{sec:aptclassi}, \textit{i.e.} steps (5-7) all fail.
	\item \textbf{Metadata N-gram Search:} APTClass considers a n-gram word search on the metadata. Due to the likelihood of duplication within the OSINT, APTClass also includes any metadata of the same file. APTClass considers all possible word n-grams of the metadata. The logic for this is the metadata is likely to include the original filename and any keywords attached by the author of the report. We also include the file path as part of the metedata as the analyst is likely to store the reports in the most relevant folder and therefore using previous file paths increases our chance of matching the right label.
	\item \textbf{N-gram Keyword Search:} APTClass extends the n-gram search to additionally include the extracted text. APTClass performs the match based on all possible n-grams of the top five keywords extracted. We empirically verified in most cases the correct label lies among the top words. APTClass uses the top five keywords but this can be increased until APTClass achieves a exact match with a corresponding linear increase in processing time. %
	\item \textbf{Remaining Text:} If APTClass fails to identify a label for a hash in steps (4-6) then it stores the text and repeats steps (3-7) using this remaining text. If it fails again then the label will be a dictionary consisting of the top five keywords from the full text and the keywords from the metadata. 	 
\end{enumerate}

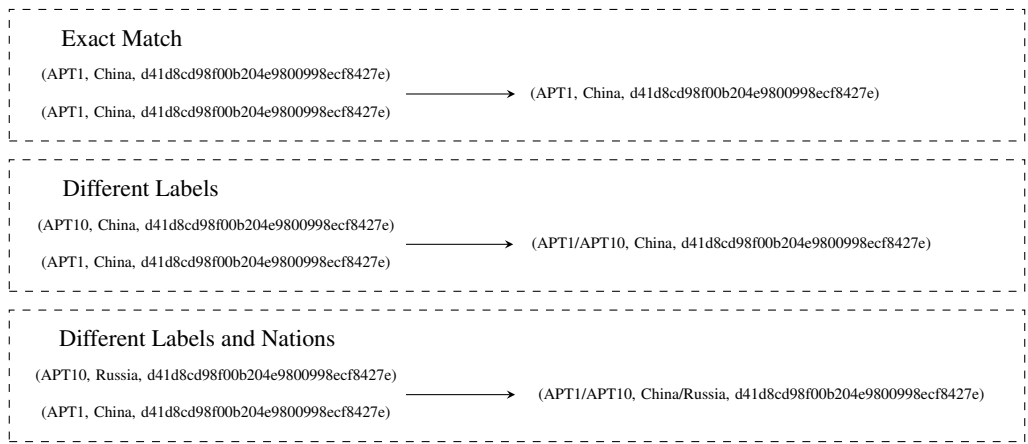
\begin{figure}[ht!]
	\begin{minipage}[t][6cm]{\columnwidth}
		\begin{center}
			\begin{tikzpicture}[node distance = 0mm]
				\tikzstyle{label} = [rectangle, minimum width=4cm, minimum height=0.5cm,text centered, draw=white, fill=white, font=\footnotesize, inner sep=0.5pt, text width=5cm, font=\tiny]
				\tikzstyle{arrow} = [thin,->,>=stealth]
				\node (BG1) at (0,0.5) [draw=black, dashed,minimum width=13.5cm,minimum height=1.75cm] {};
				\node at (-5.25,1)[font=\small]{Exact Match};
				\node (l0) [label] at (-4,0.25){};
				\node (l1) [label] at (-4,0.5){(APT1, China, d41d8cd98f00b204e9800998ecf8427e)};
				\node (l2) [label] at (-4,0){(APT1, China, d41d8cd98f00b204e9800998ecf8427e)};
				\node (l3) [label, right of=l0, xshift=6.5cm] {(APT1, China, d41d8cd98f00b204e9800998ecf8427e)};									
				\draw [arrow] (l0) -- (l3);
				\node (BG1) at (0,-1.5) [draw=black, dashed,minimum width=13.5cm,minimum height=1.75cm] {};
				\node at (-5,-1)[font=\small]{Different Labels};
				\node (l4) [label] at (-4,-1.75){};
				\node (l5) [label] at (-4,-2){(APT1, China, d41d8cd98f00b204e9800998ecf8427e)};
				\node (l6) [label, text width=5.25cm] at (-4,-1.5){(APT10, China, d41d8cd98f00b204e9800998ecf8427e)};
				\node (l7) [label, right of=l4, xshift=6.85cm, text width=5.75cm] {(APT1/APT10, China, d41d8cd98f00b204e9800998ecf8427e)};									
				\draw [arrow] (l4) -- (l7);
				\node (BG1) at (0,-3.5) [draw=black, dashed,minimum width=13.5cm,minimum height=1.75cm] {};
				\node at (-4.25,-3)[font=\small]{Different Labels and Nations};
				\node (l8) [label] at (-4,-3.75){};
				\node (l9) [label] at (-4,-4){(APT1, China, d41d8cd98f00b204e9800998ecf8427e)};
				\node (l10) [label, text width=5.25cm] at (-4,-3.5){(APT10, Russia, d41d8cd98f00b204e9800998ecf8427e)};
				\node (l11) [label, right of=l8, xshift=7.25cm, text width=6.5cm] {(APT1/APT10, China/Russia, d41d8cd98f00b204e9800998ecf8427e)};									
				\draw [arrow] (l8) -- (l11);
			\end{tikzpicture}			
		\end{center}
	\end{minipage}
	\caption{Example of APTClass cleaning process.}
	\label{fig:joining_samples}
\end{figure}

\subsubsection{Cleaning, verifying and filtering} 
Before checking the hashes discovered from the extraction process, APTClass cleans the data by joining identical hashes and collates any information which suggests mis-attribution. APTClass joins any samples with a exact match (\textit{i.e} the hash, the group name and group nation are identical). Next APTClass joins any samples where the hash and the nation are identical but the labels differ; in this case APTClass concatenates the labels. Finally, APTClass joins any remaining samples with identical hashes but nations and labels differ; in this case APTClass concatenates the labels and concatenates the nations. We provide an example of this cleaning process in \Cref{fig:joining_samples}. 
Once this step is complete, APTClass submits each MD5, SHA1 and SHA256 sample to VirusTotal to check the malware legitimacy, the file type and the corresponding hash values. After this, APTClass repeats the cleaning step above and joins together any labels for identical hashes. Finally, APTClass filters for executable samples. 

\subsection{Results}

We run APTClass using the sources listed in \Cref{tbl:aptclass-sources}, including 373 report files, 504 IoCs and 19 Yara Rules. %
The analysis takes approximately 116 hours\footnote{Approximately 80\% of the time taken is accounted by the \textit{cleaning, verifying and filtering} process, this is determined in the \textit{verifying process} by the rate limit of the VirusTotal API.} on a Ubuntu 16.04 Virtual Machine equipped with 16 vCPU and 16GB RAM. At the end of this process, APTClass returns a list of 15,660 labeled samples. The results are shown in \Cref{tbl:datasets_comparison}, together with a comparison of existing APT datasets. As we see from \Cref{tbl:datasets_comparison}, APTClass is comfortably larger than both \cite{aptdataset} and \cite{aptdataset2}. Unfortunately, there lacks the availability of the OSINT used within \cite{aptdataset} and \cite{aptdataset2} and so we cannot run APTClass on the same reports to see if there is any comparison. However, we believe the issues discussed in Section \ref{sec:malwaredatasets} and slight difference in goals of the three systems makes it very difficult to compare datasets in terms of the granularity within \Cref{tbl:datasets_groups_nations_hashes}. 

\begin{table*}[h!]
	\rowcolors{1}{white}{Gray}
	\centering
	\begin{threeparttable}
		\caption{Comparison of our dataset against both \cite{aptdataset} and \cite{aptdataset2}.} 
		\label{tbl:datasets_comparison}
		\begin{tabular}{p{7cm}L{1cm}L{1cm}L{2cm}}
			\toprule
			\rowcolor{white}
			& \cite{aptdataset} & \cite{aptdataset2} & \textbf{APTClass} \\ \midrule
			Total labeled Samples & 8,927\tnote{a} & 3,594\tnote{b} & \textbf{15,660} \\
			Number of groups & 88 & 12 & \textbf{164} \\
			Number of threat intelligence files processed & 821 & 33 & \textbf{896} \\ \midrule
			Total unknown samples & N/A & N/A & \textbf{7,485} \\ 
			Number of groups with 50+ samples & N/A & 11 & \textbf{37} \\
			Number of groups with 25+ samples & N/A & 12 & \textbf{54} \\ \bottomrule
		\end{tabular}
		\begin{tablenotes}
			\footnotesize
			\item[a] This includes file types other than ELF, Windows 32 EXE, Windows 16 EXE, Windows Installation file and Windows DLL.
			\item[b] \citet{aptdataset2} include information on a further 855 samples which are not on VirusTotal.
		\end{tablenotes}
	\end{threeparttable}
\end{table*}

\begin{table*}[ht!]
	\rowcolors{1}{white}{Gray}
	\centering
	\caption{The number of SHA256 hashes per Nation and APT Group.} 
	\label{tbl:datasets_groups_nations_hashes}
	\begin{tabular}{C{3.55cm}f{1cm}C{2.5cm}f{1cm}C{2.5cm}f{1cm}}
		\toprule
		\rowcolor{white}
		Nation &  &	APT Group &   &	APT Group & \\		\midrule
		China &            5,548 &		apt10 &             548 &	icefog &              90 \\
		India &             417 &			apt17 &            2462 &infy &             189 \\
		Iran &             637  &			apt27 &              85 &kimsuky &              77 \\
		Israel &            5,000 &			apt28 &             500 &lazarus &            1046 \\
		Italy &               6 &			apt29 &              93 &mirage &              75 \\
		Lebanon &              26 &			apt33 &              83 &muddywater &              63 \\
		Libyan Arab Jamahiriya &               1  &			apt37 &              77 &oceanlotus &             679 \\
		DPRK &            1,236 &			apt40 &             103 &patchwork &             282 \\
		Pakistan &               8 &			be2 &             110 &promethium &              89 \\
		Russia &            1,658&			\multirow{1}{*}{black vine} &             316 &rtm &              88 \\
		Turkey &              89  &			blackgear &             270 & \multirow{1}{*}{scarlet mimic} &              61 \\
		United States &              74			 &			blacktech &             333 &	sig17 &            4,992 \\
		Vietnam &             679  &			cleaver &             112 &silence &              65 \\
		& &			\multirow{1}{*}{comment crew} &             260 &ta505 &             171 \\
		& &			confucius &              87 &thrip &             105 \\
		& &			darkhotel &              94 &tick &              70 \\
		& &			fin7 &             181 &	\multirow{1}{*}{tropic trooper} &              59 \\
		& &	gamaredon group &        159 &turla &              86 \\
		& 					& higaisa &               53 &	 & \\ 
		\bottomrule&& & 
	\end{tabular}
\end{table*}

APTClass creates an overall diverse dataset with 164 APT groups from which we can create a concentrated subset consisting of 37 groups with 50 or more samples. In Table \ref{tbl:datasets_groups_nations_hashes}, we provide a breakdown of the results by the 13 nations (without potential mis-attribution) and the 37 groups with 50 or more samples. Although there exists a clear disparity among the nations, this reflects the information sources and publicly known attacks. Similarly among APT groups there are certain groups where there are considerably more samples linked to them (\textit{e.g APT17 - China} and \textit{SIG17 - Israel}), which reflects the samples by nation with both \textit{China} and \textit{Israel} linked to the most amount of samples. Overall, \Cref{tbl:datasets_groups_nations_hashes} mirrors the observations made in Section \ref{sec:apt-background} and those seen in \Cref{tbl:top_apt}. In fact, this further highlights the bias towards non-Western nation sponsored APT groups. Interestingly, only two APT groups (\textit{Oil Rig} and \textit{Emissary Panda}) of the 2020 top ten are not included in \Cref{tbl:top_apt}. Additionally, the group \textit{kimsuky} is linked to 77 samples compared to zero in \Cref{tbl:top_apt}. In general, the number of samples vary considerably to \Cref{tbl:top_apt} which is most likely because not every threat intelligence company shares their intelligence with MITRE.

\subsection{Discussion}

Even though we focus purely on extracting malware hashes from OSINT, APTClass can be enriched by extracting other indicators or relevant information from OSINT such as \textit{Tactics, Techniques and Procedures} and \textit{malware families} to build further datasets for wider research into malware analysis. APTClass also allows the user to select the sources used for the creation of the APT label list (Section \ref{sec:aptclassi}) and OSINT collection (Section \ref{sec:aptclassii}).

\begin{figure}[ht!]
	\centering
	\begin{subfigure}{0.42\textwidth}
		\centering
		\includegraphics[width=0.99\linewidth, height=5cm]{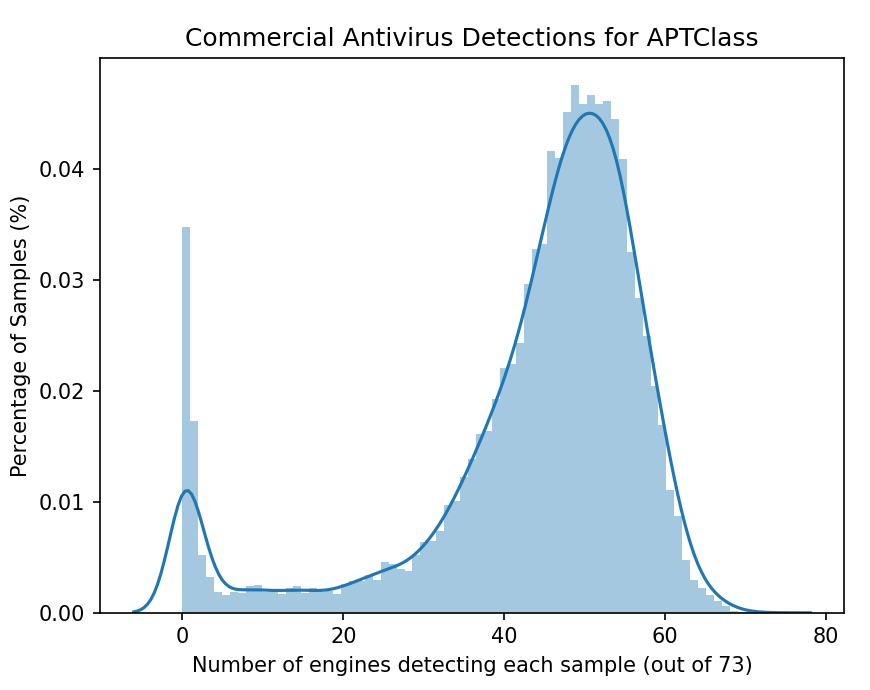}
		\caption{APTClass}
		\label{fig:malicious}
	\end{subfigure}
	\begin{subfigure}{0.57\textwidth}
		\centering
		\includegraphics[width=0.99\linewidth, height=4.75cm]{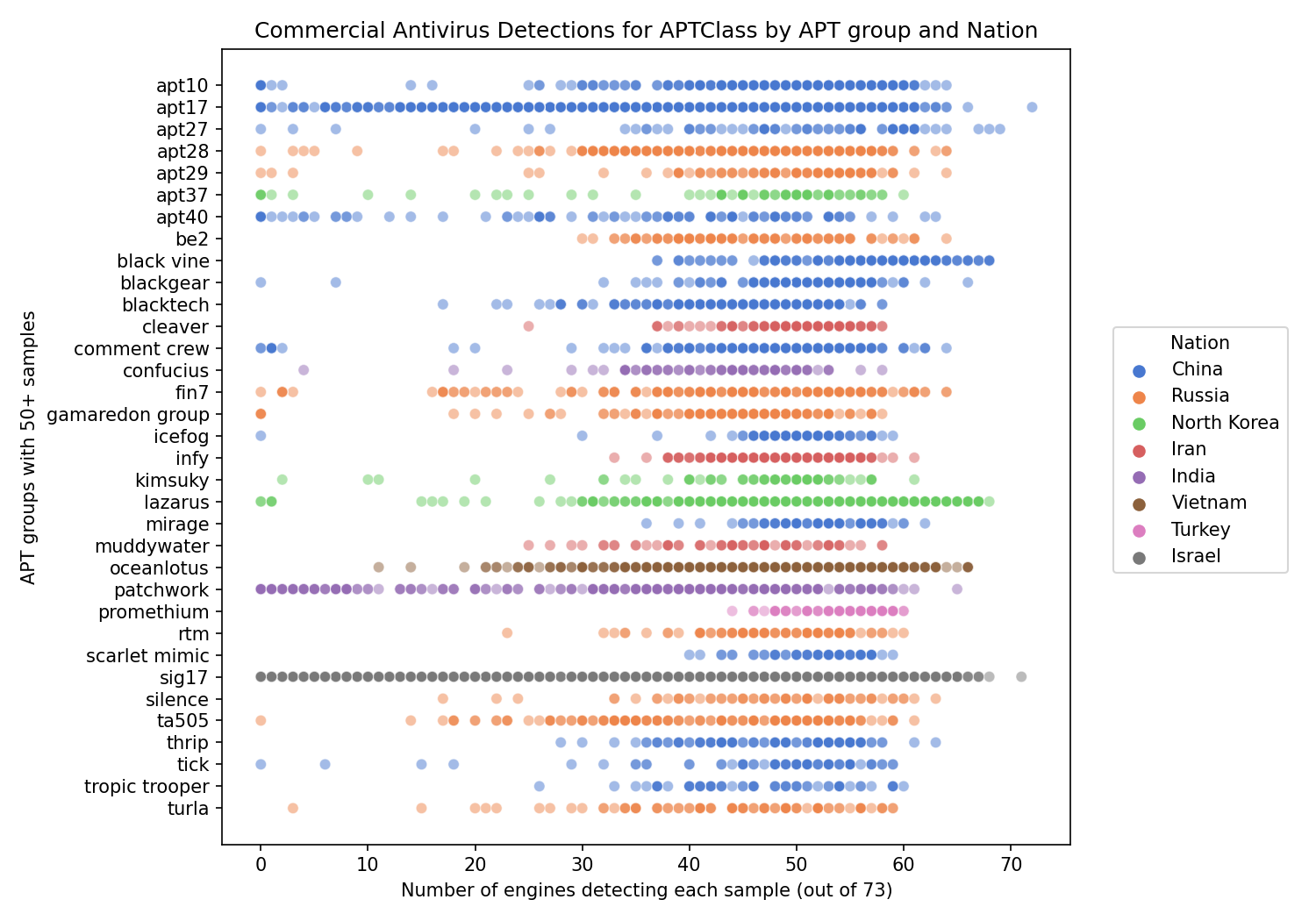}
		\caption{APT groups with 50+ samples}
		\label{fig:topapt}
	\end{subfigure}
	\caption{The detection results against up to 73 commercial anti-virus engines.}
	\label{fig:vtresults}
\end{figure}

One additional use of APTClass is it can help produce new methods for malware detection. APTClass offers a different perspective on traditional detection methods as well as testing them on the most sophisticated malicious techniques. We show this by including the detection results of APTClass against up to 73 commercial anti-virus engines from VirusTotal in \Cref{fig:vtresults}. In \Cref{fig:malicious}, we see there exists a small proportionate of APT malware which no anti-virus engines detect. Interestingly, this issues is not specific to one APT group or Nation (\Cref{fig:topapt}). These graphs highlight an unsolved problem within malware detection. Furthermore, APTClass offers a unique niche dataset for testing data modeling techniques used in the malware domain. Specifically, we can see APTClass being used to further develop and understand sophisticated adversarial attacks.

Due to the cross-domain benefits APTClass provides, we publish the code and dataset for this joint project at \url{https://s3lab.isg.rhul.ac.uk/aptclass}. We additionally welcome contributions towards evolving APTClass to continually support the research community.

\section{Conclusion}

We presented a comprehensive survey of the Malware Authorship Attribution problem by focusing on threat actor style and adversarial techniques to the current state-of-the-art systems. We specifically examine the current data modeling techniques, datasets and features used for malware authorship style. We compared the results of eighteen binary attribution systems and identified the current limitations of state-of-the-art techniques. Surprisingly, we found most of these limitations apply to all of the eighteen systems, which shows a lack of progression. Therefore, we envision our work as a source of stimulation for future research, especially for new practitioners. Furthermore, we mitigated the issue of lack of author labeled malware dataset by creating a verified dataset containing 15,660 APT samples linked to 164 APT group names and 13 nations. This is the largest dataset of this type publicly available, and can be used by researchers and practitioners as a common ground to test and compare their approaches.

\bibliography{arxiv}

\end{document}